\documentclass[a4paper, 10pt, conference]{ieeeconf}      

\IEEEoverridecommandlockouts                              

\usepackage{graphics} 
\usepackage{epsfig} 
\usepackage{float} 

\usepackage{mathptmx} 
\usepackage{times} 
\usepackage{amsmath} 
\usepackage{amssymb}  

\usepackage{dsfont}
\usepackage{soul}
\usepackage[normalem]{ulem}
\usepackage[utf8]{inputenc}
\usepackage[small]{caption}
\usepackage{graphicx}

\usepackage{amsmath}
\usepackage{booktabs}
\usepackage{algorithm}
\usepackage[noend]{algorithmic}
\usepackage{amsfonts}
\usepackage{blindtext}
\usepackage{subfig}
\graphicspath{{Figures//}}
\DeclareGraphicsExtensions{.eps,.pdf,.png,.jpg,.tif,.tiff,.ps}

\usepackage{caption}

\usepackage{outlines}

\usepackage{breqn}

\makeatletter
\let\NAT@parse\undefined
\makeatother
\usepackage[hidelinks]{hyperref}  
\usepackage{url}
\usepackage[nameinlink,capitalize]{cleveref}
\usepackage{soul}
\usepackage[colorinlistoftodos]{todonotes}

\definecolor{navy}{rgb}{0.1, 0.1, 0.8}
\definecolor{gray}{rgb}{0.6, 0.6, 0.6}
\definecolor{myblue}{rgb}{.8, .8, 1}
\definecolor{olive}{rgb}{0.1, 0.5, 0.1}
\definecolor{magenta}{rgb}{0.55, 0.0, 0.55}
\usepackage{adjustbox}
\usepackage{textcomp}
\usepackage{siunitx}
\usepackage{subfig}

\usepackage{lipsum}

\usepackage{xspace}

\newcommand{\citet}[1]{\citeauthor{#1}~\shortcite{#1}}
\newcommand{\citep}{\cite}

\newcommand{\titlename}{How will electric vehicles affect traffic congestion and energy consumption: an integrated modelling approach}

\title{\LARGE \bf \titlename}

\author{Artur Grigorev$^{1}$, Tuo Mao$^{1}$, Adam Berry$^{1}$, Joachim Tan$^{2}$, Loki Purushothaman$^{1}$, Adriana-Simona Mihaita$^{1}$
\thanks{$^{1}$ Most authors are with the University of Technology in Sydney, Faculty of Engineering and IT, School of Computer Science: 61 Broadway Ultimo, NSW, Australia. Corresponding authors contact: {\tt\small artur.grigorev@student.uts.edu.au, adriana-simona.mihaita@uts.edu.au}}
\thanks{$^{2}$ This work was sponsored by AEMO the Australian Energy Market Operator. The authors are highly grateful for the guidance and support of Joachim Tan, Adrian Grantham, Greg Staib and Andrew Turley. Joachim Tan is a Principal Manager in the Australian Energy Market Operators and main lead of the UTS AEMO project collaboration.} 
\thanks{$^{3}$ The authors are highly grateful for the guidance and data support of Transport for NSW, Australia, more specifically Virginie Vernin, Bill Chen, Blake Xu and Marwan Daizli.} 
}
\begin{document}

\maketitle
\thispagestyle{empty}
\pagestyle{empty}

\begin{abstract}
This paper explores the impact of electric vehicles (EVs) on traffic congestion and energy consumption by proposing an integrated bi-level framework comprising of: a) a dynamic micro-scale traffic simulation suitable for modelling current and hypothetical traffic and charging demand scenarios and b) a queue model for capturing the impact of fast charging station use, informed by traffic flows, travel distances, availability of charging infrastructure and estimated vehicle battery state of charge. To the best of our knowledge, this paper represents the first integrated analysis of potential traffic congestion and energy infrastructure impacts linked to EV uptake, based on real traffic flows and the placement and design of existing fast-charging infrastructure. Results showcase that the integrated queue-energy-transport modelling framework can predict correctly the limitations of the EV infrastructure as well as the traffic congestion evolution. The modelling approach identifies concrete pain points to be addressed in both traffic and energy management and planning. The code for this project can be found at : \textcolor{blue}{\url{https://github.com/Future-Mobility-Lab/EV-charging-impact}}   
\end{abstract}



\section{Motivation} 
\nocite{appendix}
The adoption of electric vehicles (EVs) at large scale around the globe is largely dependent on two key factors: charging accessibility for its consumers and availability of sufficient electrical grid capacity to support vehicle charging, particularly where fast charging infrastructure is necessary. Under ideal conditions for the consumer, access to an EV charging station should be just as convenient as current access to a petrol station. Several studies have been conducted in recent years to model and identify suitable locations for such charging stations \cite{CSONKA2017768}, \cite{wevj10010012}. The majority of these studies are based on the behavioural pattern of combustion engine (CE) drivers. However, the behaviour of electric vehicle (EV) drivers is significantly different in terms of location selection and their waiting time to charge \cite{CSONKA2017768}, \cite{CHAKRABORTY2019255}. Multiple factors can influence the decision of EV drivers to choose a specific location to charge, among which we cite: the travel time towards the closest EV station, the state of charge of the battery (SoC) while driving, the waiting time in queue at the EV station, the total charging time that a driver needs to plan different activities around while the car is charging. The stochasticity of this charging behaviour makes the modelling of EVs a hard problem to solve, which may delay evidence-based infrastructure planning at a government level and reduce investment confidence for commercial infrastructure and electric vehicle businesses. 
Therefore, the inability to charge conveniently in the traffic network coupled with limited charging stops makes the problem challenging from both operational and planning perspectives. Even more challenging, evaluating the grid impact of EV adoption comes along not only with a consumer random behaviour affected as well by traffic congestion variation, but with the physical constraints of both infrastructure availability and  maximal energy capacity of each serviced area.     

These challenges represent the motivation of this study which to the best of our knowledge is the first to propose an integrated bi-level framework comprised of a dynamic traffic simulation modelling coupled with queueing behaviour modelling for estimating the impact of traffic demand and congestion on EV fast charging behaviour and vice-versa.

\subsection{Related works}\label{Literature_review}

\begin{figure*}[ht]
	\centering
	\includegraphics[width=1\textwidth]{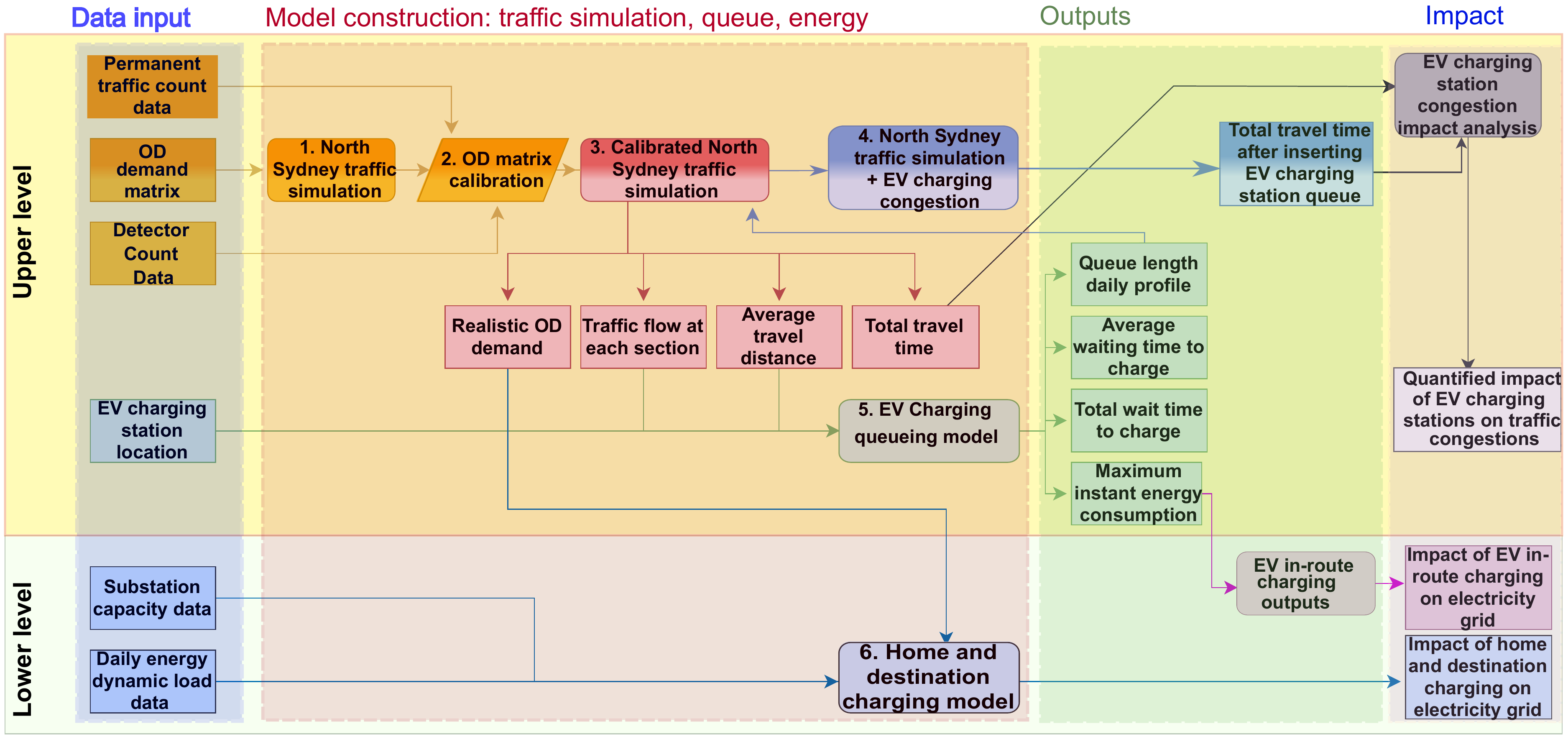}
	\caption{Bi-level framework of transport and energy modelling containing: A) Upper Level focused on a hybrid transport and queue modelling approach for quantifying EV charging on traffic congestion and vice-versa and B) Lower level focus on building a home-and-destination charging model using substation capacity data and the daily and dynamic energy load data in the study area.}
	\label{Fig3_framework}
\end{figure*}

Several studies \citep{MOON201864}-\citep{Qian2011},  have been conducted to forecast the electricity demand that will be created by the penetration of EVs into the market in the future. There are also works to determine and forecast the impact this electricity demand will have on the existing power grids \citep{DULAU2020370}. However, the majority of these studies either orient themselves towards analysing the impact of the electricity consumption on the grid using hypothetical assumptions of the charging locations or very broad estimations of the major attraction points in specific areas (p-median or location-allocation problem) such as shopping malls, schools, hospitals and workplaces \citep{Ghamami2016}. 
Alternatively, some studies that have considered the user equilibrium (UE) approach to optimise the route choice so that the travel time is minimized when recharging \citep{HE2014306}, \citep{XU2017138}. However, these studies have several assumptions such as that congestion will not affect the electricity consumption or that the path lengths are within battery driving range. However, congestion represents an important factor to be taken into consideration in the EV impact modelling, especially when the fast charging stations reach their plug-in capacity. 
Some studies have developed queuing models that take into consideration the station capacity, the charging time and EV waiting time for example \citep{Zhu2017PlanningOE}; others have considered as well the maximal charging demand which satisfies the user's maximal tolerable queuing time \citep{Lu2015} or have proposed a global queue system that would dispatch EVs to available charging stations \cite{en10070952}. While attention was given to parameters describing the charging situations inside the EV stations, studies did not take into consideration for example the battery state of charge (SoC) when routing decisions are being made, nor the travel time affected by increased demand. Other queuing approaches can be found in \citep{XIAO2020101317} where they took consideration of the fact that charging stations can have a different number of plugs with various charging capabilities.   

In \citep{JUNG2014123} there is a tentative approach to couple together simulation and optimisation techniques for optimising a taxi fleet and similar approaches have been proposed which consider for example the planning of routes according to an activity-based model \citep{USMAN2020745} or agent-based simulation \citep{KLEIN2020102475}. We believe, however, that there remains a key gap in the use of traffic simulation models to explore the interrelationship between charging activity, traffic behaviour and road congestion outside of strictly fleet-based settings. Similarly, embedding EV queue modelling at each station in the simulation model remains an under-explored research topic.

From the above discussions, we identify several gaps in the current literature. Firstly, several studies do not take into consideration future vehicle demand increases and how this may impact EV fast charging accessibility, queuing times, traffic volumes and emergent congestion related to the use of fast charging infrastructure. The inter-dependency between congestion, demand and queue spill-back from EV charging stations should be modelled together for both current and future traffic scenarios. 
Secondly, the number of available plug-in slots and their charging rates should be embedded in the queue model and constraints should be applied around realistic charging queue lengths; this is to avoid assumptions of infinite queue models which can accept all incoming EV requests. 
Thirdly, several studies propose an optimisation of charging stations based on assumption that people charge either at home or at their destination, whereas in reality, the charging behaviour is dynamic and can include mid-route charging stops as well based on battery SoC, charging availability and convenience (i.e. charging at a shopping centre); this is a factor that would need to be embedded in any queue modelling approach. Finally, analysis of electric vehicle behaviour should include the electrical grid impact, as the shape and extent of charging demand will inevitably impact electrical network infrastructure planning and operation, particularly where charging demand is aligned with peak electrical demand.

\subsection{Paper Contributions}\label{Contributions}

In our work we aim to address previous gaps by proposing an integrated framework of traffic modelling with stochastic queue behaviour (the upper-level which can be used for en-route charging behavioural modelling showcased in this paper) and a data-driven energy consumption estimation (the lower-level for home and work charging estimation which we will present in a future paper extension). The approach has been applied on the Northern Sydney urban area and is built using realistic data sets in terms of: traffic demand, vehicle counting, EV charging stations (locations, number of plugs, type of plugs, etc.), electricity zone substation capacity data and daily electricity load profiles for those substations (sufficient for determining peak electricity demand times).  

Firstly, we propose an integrated traffic simulation and queue behavioural model for capturing not only the dynamic nature of traffic congestion but also the queue impact around existing EV charging stops. The microscopic traffic simulation is calibrated against real-world traffic flow counts and is used to evaluate traffic congestion under different EV and charging station settings. We also explore the potential impact of hypothetical future demand scenarios. The queue model which works on a first-in-first-out approach is tailored for each station and ingests several simulation outputs (travel time to reach the station, calibrated flow, travel distance, calibrated OD demand), as well as the SoC of the batteries, the number of available charging plugs and their charging rates. We denote this model as $EV-Q(M/M/n_{ep}^{S_i}/N_c^{S_i})$ where $N_c^{S_i}$ is the maximum number of car slots available at each station (for both waiting and charging) and $n_{ep}^{S_i}$ is the number of plugs (chargers) for each station. The EV-Q model generates queue length estimations which are then fed to the micro traffic simulation model for congestion impact analysis.  Secondly, we explore the behaviour based on both observed road demand from 2016 and two hypothetical scenarios: one with a $15\%$ increase in road use and another with a $30\%$ increase. Across all scenarios, we explore impacts of different levels of assumed EV penetration. Lastly, for all simulations, we estimate total electrical demand associated with fast charging, enabling a review of potential electrical infrastructure impacts.


The paper is organised as follows: \cref{section_methodology} introduces the bi-level framework for our integrated queue, energy and transport simulation modelling, \cref{Case_study} presents a case study evaluation on the Northern Sydney area including traffic simulation modelling (construction, calibration and validation), queue modelling (construction and validation) and EV impact evaluation across different penetration rates and demand scenarios. The last section is reserved for conclusions, limitations of this work and further directions.

\section{Methodology}\label{section_methodology}
\begin{figure}[ht]
	\centering
	\includegraphics[width=0.49\textwidth]{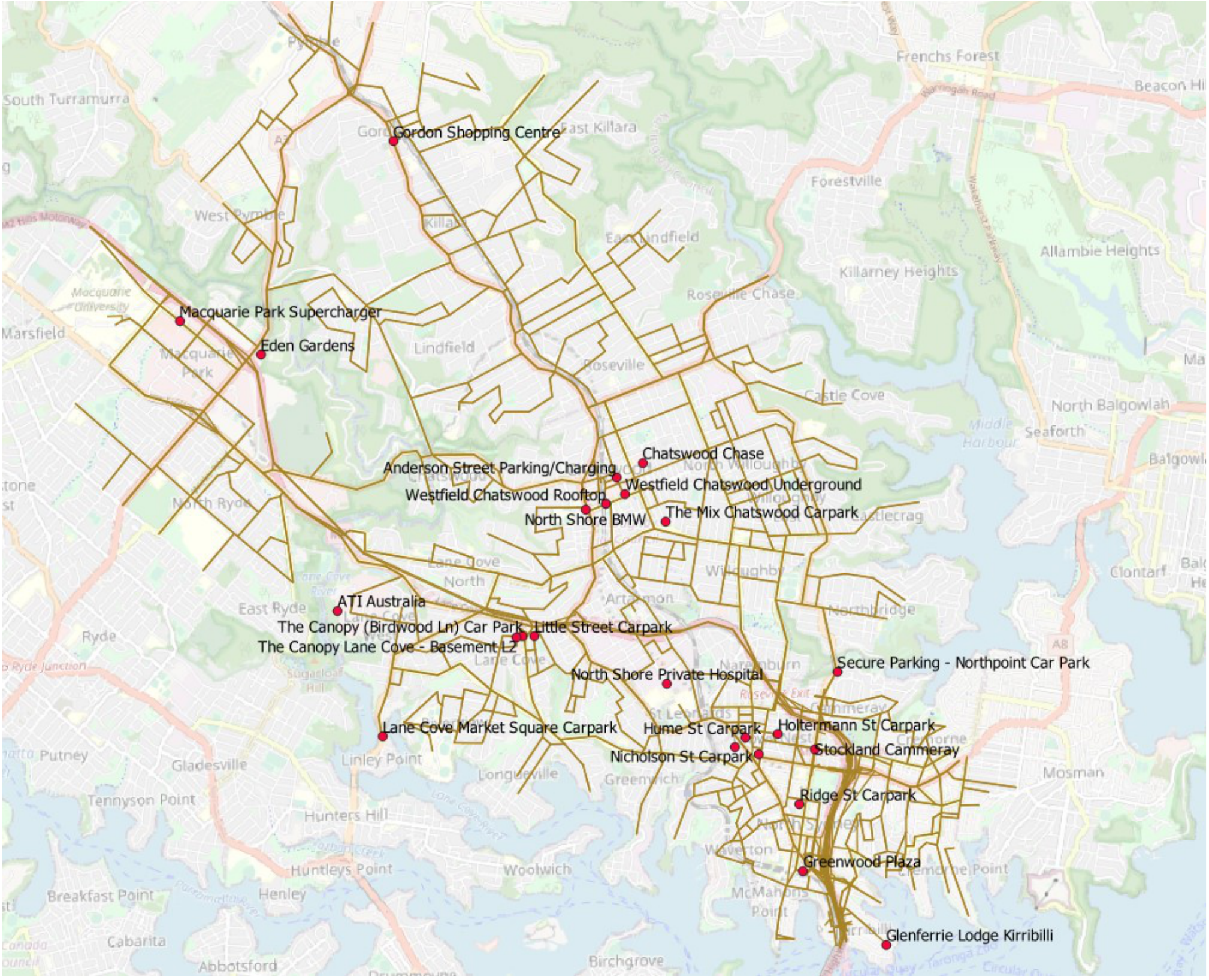}
	\caption{Network layout of the North Sydney traffic simulation model (NSTM) including all existent electric vehicles charging stations (marked as red).}
	\label{Fig2_Sydney_train_network}
\end{figure}

The bi-level framework we propose in this paper is illustrated in \cref{Fig3_framework} and is comprised of two major parts: A) the Upper level construction with the purpose of building a hybrid model for analysing the quantified impact of EV charging station availability on traffic congestion and vice-versa, and B) the Lower level construction which focuses on building a home-and-destination charging model using substation capacity data and the daily and dynamic energy load data in the study area. Due to lack of space the current paper will only include the upper level modelling and selected results. Each model comprises of several modules coloured differently, with dedicated entry data streams and output measures, summarized here below.

 \begin{figure*}[ht]
	\centering
	\includegraphics[width=1\textwidth]{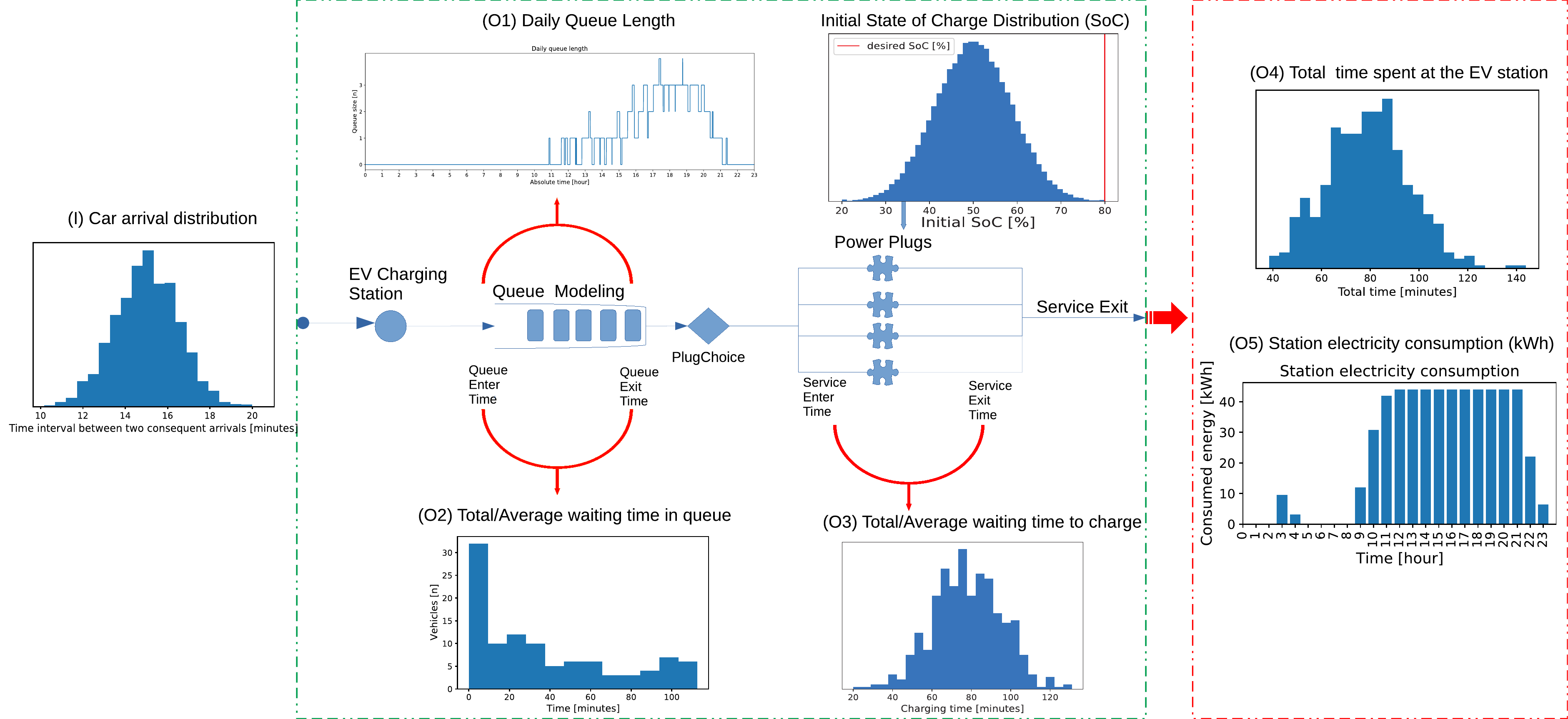}
	\caption{EV charging representation using a queue modelling approach of vehicles arriving at each charging stations.}
	\label{EV_Q_model}
\end{figure*}

\subsection{Transport simulation modelling}\label{simulation}

In order to evaluate the impact of current and future traffic demand on the electric vehicle fast charging, one needs to be able to evaluate realistic travel distances, travel times and traffic flows for the road network under various dynamic traffic conditions. The traffic simulation we are building is a microscopic dynamic model which is comprised of all vehicle movements, from any internal origins to their destinations and intermediary charging stops along the route. 

We select North Sydney as an application area (see \cref{Fig2_Sydney_train_network}) and refer to our simulation model as NSTM. All notations are provided in \cref{Tab_notation_initial}. The area has been selected due to its large business hubs, major train stations, new metro line and major commuting trips to/from Sydney CBD on a daily basis. The NSTM model consumes in its input: a) the network layout with real road structure definition, b) OD demand matrices ($OD(T_r)$), c) traffic signal control definitions and d) the placement and characteristics of real-world EV charging stations deployed in Northern Sydney ($S_i, i\in \{1,..25\}$). Overall the simulation model contains the daily trips of almost 100,000 people (according to 2016 Australian Census data), 25 EV charging stations, 41 permanent traffic count stations, 240 SCATS controlled intersections, 1,919 traffic flow detectors and 2,750 road sections. 

The micro-simulation model goes through several procedures of initialisation, calibration and validation including: 
i) an initial static traffic assignment (STA) procedure of the original traffic demand to available routes, 
ii) a static OD adjustment (SOD) using permanent traffic flow counts, 
iii) a daily demand profiling for AM and PM peaks, 
iv) a Dynamic OD adjustment (DOD) using detector flow counts on an hourly basis for 24-hour time interval which helps us to calibrate the initial demand matrix, 
v) a dynamic user equilibrium (DUE) assignment which is validated against both permanent and detector flow counts, and finally 
vi) a microscopic simulation run consuming on input as well the estimated queue lengths on each road section in the vicinity of the EV charging stops (this is an output of the EV-Q model described in the next sub-section).   

The major outputs of the micro NSTM model are the following: 
a) a calibrated and more realistic OD demand matrix $OD_c(T_r)$, 
b) the total travel times for AM/PM peaks, 
c) the calibrated traffic flows and finally 
d) the average travel distance (the last two used as inputs to the EV-Q charging model for each station). 

The NSTM is used for testing the impact of various scenarios: a baseline scenario based on observed traffic demand for 2016, and two hypothetical growth scenarios, with a $15\%$ and $30\%$ increase in total demand, respectively. The scenarios allow us to explore how changes in traffic demand interact with electric vehicle penetration rates and EV infrastructure availability to effect road use, queuing times, congestion and electricity demand.

\subsection{Queue modelling of EV charging} \label{queuemodelling}
While the traffic simulation model has the objective of providing dynamic insights on the traffic congestion in the neighbourhood, it does not have the capability of replicating the queueing behaviour of drivers that need EV charging. Therefore, by using simulation outputs such as car average travel distance and traffic flow on road sections, we develop a queue model to quantify the impact of the waiting behaviour around existing EV stations and feed this behaviour back to the simulation model. 
The traffic simulation and the queue models are connected to each other and run in parallel consuming each-other outputs. 
\cref{EV_Q_model} presents the overall queue modelling framework (EV-Q) which we further detail. 

\begin{table}[ht]
\begin{center}
\caption{Notations in use for EV modelling.\label{Tab_notation_initial}}
\begin{adjustbox}{width=0.49\textwidth}
\begin{tabular}{l|l}

      \textbf{Variable} & \textbf{Definition}\\
      \hline
       $NSTM$ & North Sydney Traffic Model, \\ 
       $NSTM_c$ & North Sydney Traffic Model calibrated, \\ 
	   $T_r$ & the 1h time interval during a 24-hour period \\
      & where $r\in \{1...24\}$;\\
      	  $ OD(T_r)$ & the Origin-Destination matrix containing the \\
      & number of assigned car trips in the network in time period $T_r$\\
       $ OD_c(T_r)$ & the calibrated Origin-Destination matrix\\
       $Tf_k(T_r)$ & the traffic flow for each road section $k$ in the network \\
       & during time interval $T_r$\\
        $Avg_d(T_r)$ & the average travel distance of cars during $T_r$\\

        $TT_{tot}^{e}(T_r)$ & the car total time during $T_r$ and specific $EV_p$ \\

		\hline 
      $N$ & the total number of EV stations in the network, \\ 
      $S_i, i\in \{1,..,N\}$ & the EV station ID,\\
      $N_c^{S_i}$ & the maximal number of cars inside a station \\ 
      $EV_p$ & the EV penetration rate in the network  \\
      	  $\lambda^{S_i}$ & mean-arrival time of cars at an EV station \\
      	  $n_{ep}^{S_i}$               & number of electric plugs per station \\
      $B_c(T_r) [kWh]$ & battery size of arriving cars during $T_r$\\
$SoC^{S_i}$ & the initial state of charge for vehicles arriving at station $S_i$\\
      $\rho^{S_i}$ & the service density of a station $S_i$\\

      $\overline{Q^{S_i}}(T_r)$ &  (O1) Mean queue length of an EV station $S_i$ per hour\\
      
 	  $\overline{WT_q^{S_i}}(T_r)$ & (O2) Mean waiting time in queue at an EV station $S_i$\\

	   $\mu_t^{S_i}(T_r)$ & (O3) Mean service time to charge at an EV station $S_i$\\
      
	   $WT^{S_i}(T_r)$ & (O4) Total time spent overall at an EV station $S_i$\\
       $E^{S_i}(T_r)$ & (O5) Total energy consumption of an EV station $S_i$\\

           $MaxQ^{S_i}(T_r)$ & (O6) Maximum recorded queue length of an EV station $S_i$\\
      	   $MaxW_q^{S_i}(T_r)$        & (O7) Maximum waiting time in queue at an EV station $S_i$\\
	        	   $MaxWT^{S_i}(T_r)$ & (O8) Maximum time spent overall at an EV station $S_i$\\
           $MaxE^{S_i}(T_r)$ & (O9) Maximal energy consumption of an EV station $S_i$\\

		\hline
            
%
%
%

    \end{tabular}
      \end{adjustbox}
  \end{center}
\vspace{-8mm}

\end{table}
Let $N_c(T_r)$ be the total number of cars in the traffic network during time interval $(T_r)\in \{ 0,1,2,...23h\}$, $EV_p$ the percentage of vehicles that will need to charge and $S$ the total number of EV charging stations (25 in our study according to the real world set-up \citep{EVmapAU}). Each EV station has a specific number of plugs (chargers) which we denote $n_{ep}^{S_i}$  and which follows a real set-up arrangement in our network (7 stations have only 1 plug available, 3 have 4 plugs, 2 have 3 plugs, 10 have 2 plugs, 1 has 7 plugs and only 1 has 10 plugs available). We denote the queue model as $EV-Q(M/M/n_{ep}^{S_i}/N_c^{S_i})$ where the first M stands for EV arrivals following a Gaussian distribution with mean arrival time at each station $\lambda^{S_i}$, the second M represents the distribution of the charging times with $\mu^{S_i}$ mean-service time and $N_c^{S_i}$ is the maximum number of car slots available at each station (for both waiting and charging). For this paper, we follow the queue modelling procedures from \citep{XIAO2020101317,Tian2017} which are using the Markov Chain theory for probability calculation of charging at specific plugs of a station, with the assumption that the number of waiting cars inside the stations are limited to a maximum threshold.  

The cars arrive at each EV charging station following a normal distribution with mean-inter-arrival (mIA) times denoted as $\lambda^{S_i}$ (this is calculated from real data-streams of vehicles captured by the traffic count detectors in the vicinity of each EV station in our network; mIA is calculated hourly from traffic flow with a variation of $\pm 20\%$ following normal distribution; therefore this is estimated based on simulation of traffic flow at vicinity of the charging stations). For fairness of charging, the arriving cars are being charged on a First-Come-First-Served (FCFS) approach to the next available plug in the network (statistics on emergent queues are stored in O1-\cref{EV_Q_model}). The service time (O3 in \cref{EV_Q_model}) is then calculated individually for each arriving car by taking into consideration the car State of Charge ($SoC^{S_i}$) as follows: 
\begin{equation}
  (O3): \mu^{S_i}(T_r) = \frac{SoC^{S_i}(T_r)*B_c(T_r)}{Pp^{S_i}}
\end{equation}
where $SoC^{S_i}(T_r)$ represents the state of charge for arriving cars, $B_c(T_r)$ is the battery size and $Pp^{S_i}$ is the charging rate of the fast charger of a station $S_i$ (which follows the real EV station information with the specification that majority have a 22kW capability \citep{EVmapAU}). We assume that the initial State-of-Charge will follow a normal distribution between $20\%-80\%$ according to efficient Li-Ion battery use \cite{Qcastano2015dynamical}. 
Meanwhile the battery size can take several values which have been used for sensitivity and robustness analysis: $B_c(T_r) [kWh] \in \{50,58,66,74,82,90,98,106\}$, according to a 10\% difference in battery size compared to Tesla Model 3 (Standard Range) \cite{Q82}. Based on arrival and service times, one can then express the EV charging density $\rho^{S_i}$ as:
 \begin{equation}
  \rho^{S_i}(T_r) =  \lambda^{S_i}(T_r)/\mu^{S_i}(T_r)
 \end{equation}
Based on the above formulations and the Markov Chain stationary conditions regarding the matrix of transition probabilities $P$ and the stationary distribution ($\vec{\pi}\cdot P = 0$), we can first calculate the stationary distribution (the probability that a vehicle will charge at a specific plug in the EV station) in a recursive form as follows:
\begin{equation}
	\pi_k^{S_i} = \left\{
					\begin{array}{ll}
                     \frac{1}{k!}\left(\rho^{S_i}(T_r)\right)^{k}\pi_0^{S_i}    & \mbox{if } \ k \leq {n_{ep}^{S_i}} \\
                 \left(\frac{1}{n_{ep}^{S_i}}\right)^{(k-n_{ep}^{S_i})} \cdot \frac{1}{{n_{ep}^{S_i}}!}   {\left(\rho^{S_i}(T_r)\right)^{{n_{ep}^{S_i}}k}} \pi_0^{S_i}   & \mbox{if } \ k \geq {n_{ep}^{S_i}}
    \end{array} 
	\right.
\end{equation}
By further replacing in the total probability condition ($\sum_{j} \pi_j = 1$), one can easily calculate the $\pi_0$ probability that no EV will charge at a specific station $S_i$; Implicitly the average queue length (O1 in \cref{EV_Q_model}) and the average waiting time (O2 in \cref{EV_Q_model}) can then be estimated based on the surplus of vehicles waiting to charge, the capacity of a charging station and the probability of these vehicles to arrive at a station $S_i$. 

\begin{align}
(O1):Q^{S_i} (T_r)              & = \sum_{k=n_{ep}^{S_i}}^{N_c^{S_i}}(k-n_{ep}^{S_i})\cdot \pi_k^{S_i} \\
(O2):\overline{WT_q^{S_i}}(T_r)  & = \frac{Q^{S_i} (T_r)}{ \lambda\left(1-\pi_k^{S_i}\right)}
\end{align}
Other outputs of the EV-Q model can then be calculated on an hourly basis as follows:
\begin{align}
(O4):WT^{S_i}(T_r) & = W_q^{S_i}(T_r) +  \mu^{S_i}(T_r) \\
(O5):E^{S_i}(T_r)  & = PP^{S_i} \cdot \mu^{S_i}(T_r) 
\end{align}
where $WT^{S_i}(T_r)$ represents the total waiting time to charge at a station $S_i$ (O4 in \cref{EV_Q_model}), and $E^{S_i}(T_r)$ is the total energy consumption to charge during a specific time interval at a given EV station (O5 in \cref{EV_Q_model}). Other model outputs are extracted such as minimum and maximal queue length, waiting time and energy consumption (O6-O9). The EV-Q model outputs are further used in the global traffic-energy simulation modelling framework: a) the average queue length as an input in the NSTM calibration process to realistically simulate the impact of queue waiting to charge even outside of the EV stations and b) the energy consumption estimation for the impact of fast charging on electricity distribution networks.

\section{North Sydney Case Study}\label{Case_study} 

\subsection{Simulation calibration and validation}\label{calibration}

\begin{figure}[ht]
	\centering
	\includegraphics[width=0.4\textwidth]{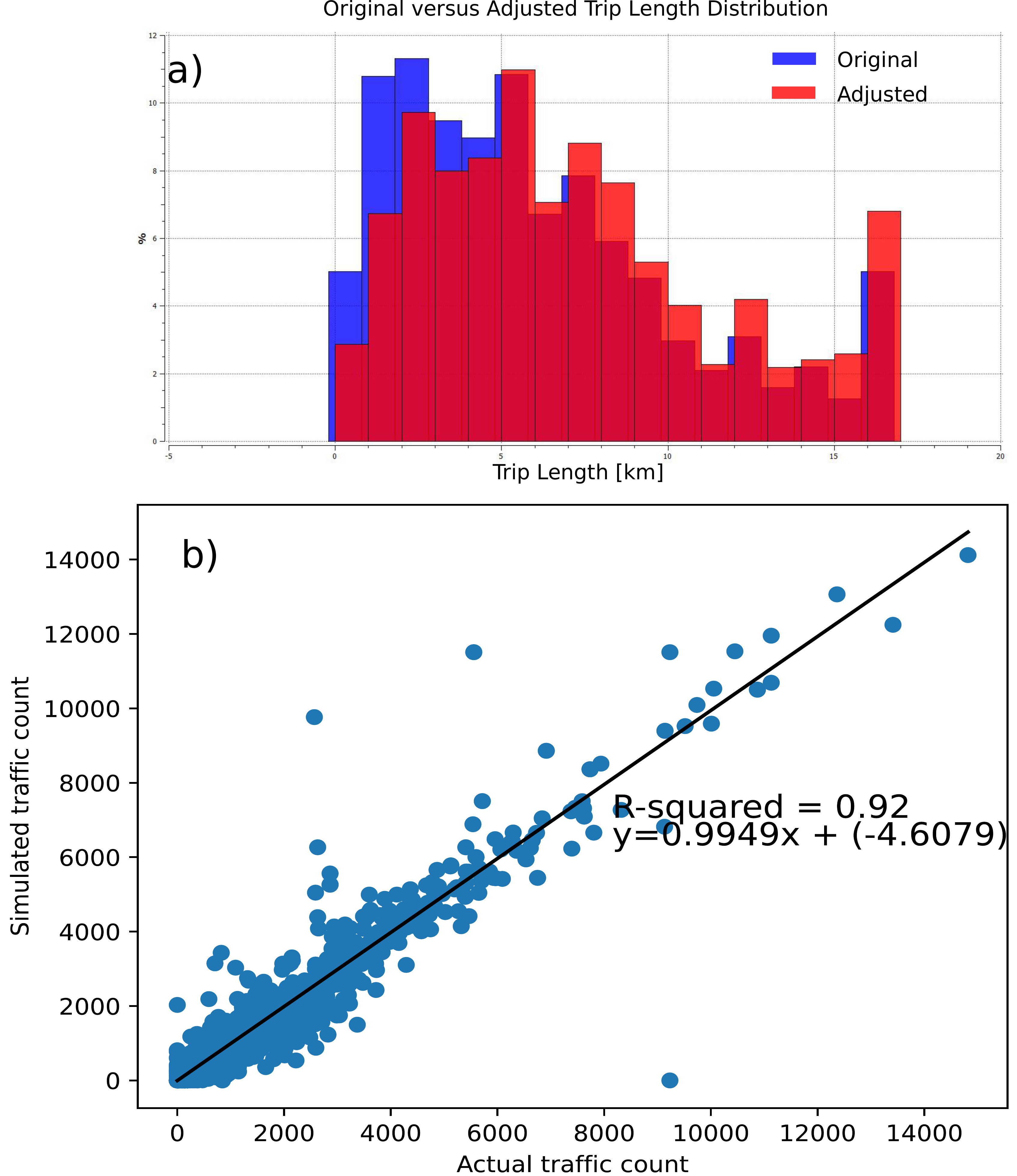}
	\caption{Validation of NSTM model for the AM peak via a) Trip Length Distribution and b) $R^2=0.92$ metric.}
	\label{AM_NSTM_RESULTS}
	\vspace{-0.3cm}
\end{figure}

After the initial construction of the NSTM model, the 200-pair OD matrix $OD(T_r)$ is calibrated by using permanent traffic flow counts from 2016 records (marked in green in \cref{Sydney_network_layout_EMME_Centroids} from the supplementary material~\cite{appendix}). The $OD(T_r)$ matrices are split to contain $93\%$ of regular car trips and $7\%$ of truck trips in the network according to real-life records. After the calibration, the dynamic and hourly demand matrices are then validated against traffic counts received from lane detectors placed at signalised intersections (showcased in \cref{Sydney_network_layout_PTC_stations} and \cref{SCATS_Control_intersection} from the supplementary material ~\cite{appendix}). The results for the AM-peak NTSM calibration and validation are provided in \cref{AM_NSTM_RESULTS}, where the Original versus Adjusted Trip Length distributions follow very similar trends with less than $2\%$ difference between them (\cref{AM_NSTM_RESULTS}a)); similarly, the $R^2$ metric calculated between simulated and the real traffic demand scored 0.92 in accuracy which validates once again the good calibration of the model (\cref{AM_NSTM_RESULTS}b)). The NSTM calibration and validation for the PM peak indicates similar performance of $R^2=0.92$ and are provided in the complementary material~\cite{appendix}-\cref{PM_validation}. We make the observation that while we validated the NSTM model against real-world traffic data from 2016, our two hypothetical growth scenarios (OD15 and OD30) cannot be similarly validated and therefore are best thought of as ``what-if'' scenarios. The outputs of the simulation model are further used as inputs in both the queue modelling and energy impact evaluation, as detailed in following section.

\subsection{Queue Model results}\label{EVQ_results}

The traffic flow $Tf_k(T_r)$ and the average travel distance $Avg_d(T_r)$ have been used as inputs in the EV-Q model with the aim of calculating the percentages of EVs needing recharging at each station, and the travel distance they had to travel before recharging. The daily fluctuations of traffic flow in the network can greatly influence the charging needs (see \cref{Traffic_flow_fluctuations_in_the_network} in~\citep{appendix}), and the number of EVs that will seek recharging based on the nearest station.  

Therefore, we apply the EV-Q model under several scenarios which allows an extensive simulation of current and potential future traffic conditions, by varying several input parameters such as the battery size of the vehicles $B_c(T_r)[kWh]$, the penetration rates of EVs that need to charge $EV_p \in \{ 0.01,  0.02, .. 0.05, 0.1, 0.12, ... 0.2, 0.21...0.3, 0.5, 1, 2, 5\} [\%]$ as well as our three traffic demand scenarios. Overall, the EV-Q modelling allows us to analyse more than 90,000 different scenarios with modeling precision down to a second-by-second. 



%

\begin{figure}[h]
 \centering
	\includegraphics[width=0.49\textwidth]{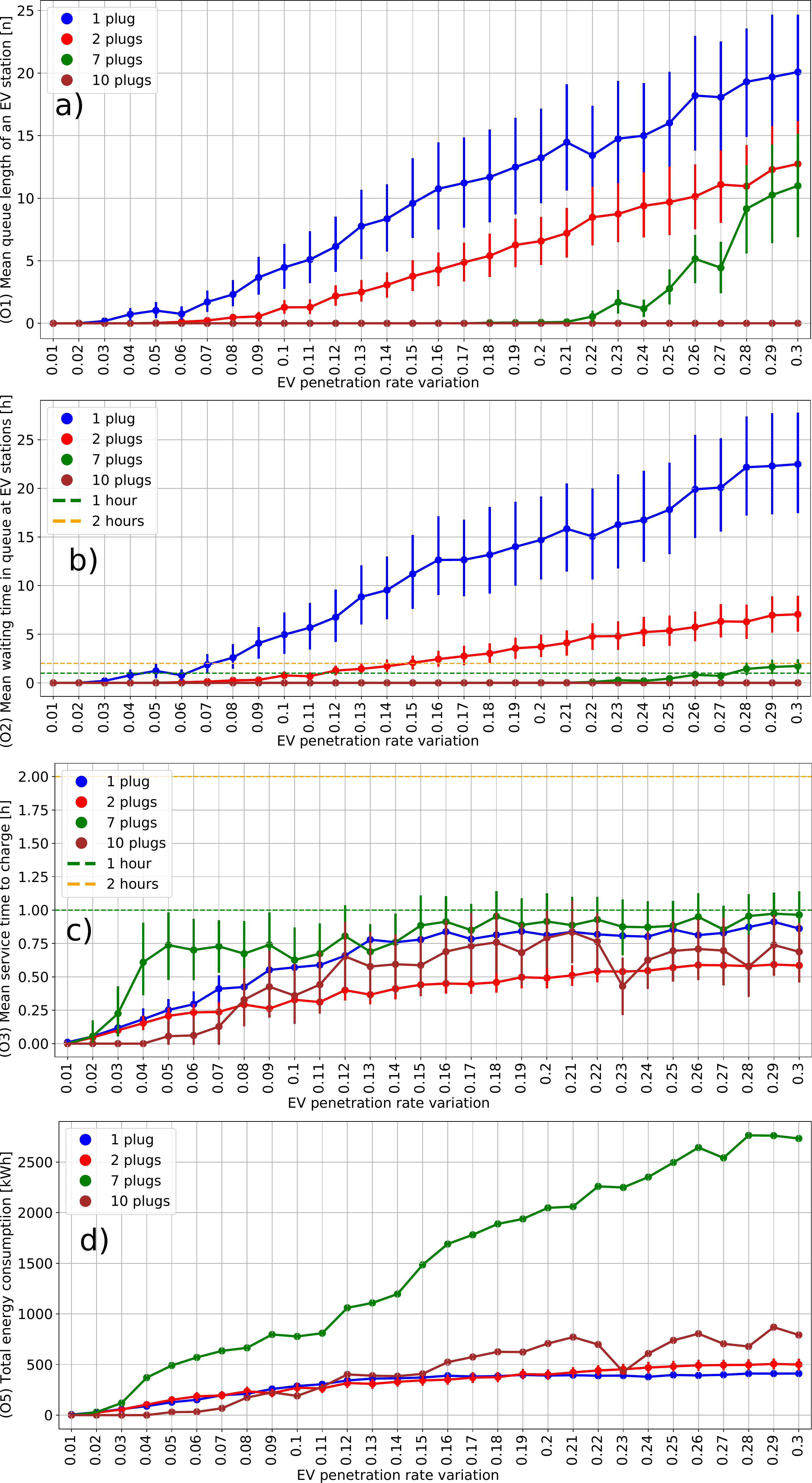}
	\caption{The impact of EV penetration rates on queue lengths, waiting times, service times and energy consumption.}
	\label{O1_O2_O3_O5_2016}
\end{figure} 

%
%
%
%
The nine outputs of the EV-Q summarised in \cref{Tab_notation_initial} are aggregated hourly and used for analysing the energy impact when drivers decide to apply an in-route charging approach (samples of O1-O5 are represented already in \cref{EV_Q_model} and several aggregations by station/area are extracted). A detailed selection of averaged outputs from the EV-Q model is provided in \cref{O1_O2_O3_O5_2016}, where we choose $B_c=82kWh$ and vary $EV_p$ between $\{0.01, 0.02, ..0.3\}$ (higher variations are presented in~\cite{appendix}-\cref{2016_2026_2036_raw_results_table}); we analyse the mean queue length (O1-\cref{O1_O2_O3_O5_2016}a)), the mean waiting time (O2-\cref{O1_O2_O3_O5_2016}b)), the mean service time (O3-\cref{O1_O2_O3_O5_2016}c)) and the mean energy consumption (O5-\cref{O1_O2_O3_O5_2016}d)). On each figure we also plot the impact of the number of plugs on the EV-Q model; for example, in \cref{O1_O2_O3_O5_2016}a), the mean queue length recorded across EV stations with one plug (blue dots and their confidence interval) indicates that cars start to queue from an $EV_p=0.03\%$ and can easily reach 20 cars if $EV_p=0.3\%$; stations with 7 charging plugs only start to accumulate cars in the queue from $EV_p=0.21\%$ while reaching a mean of 10 cars in the queue for $EV_p=0.3\%$; the best performing stations are the ones with 10 charging plugs which present zero queue accumulation until $EV_p=1\%$ (see~\cite{appendix}-\cref{Queue_for_10_plug_stations}). The mean waiting time at stations (\cref{O1_O2_O3_O5_2016}b)) is highly influenced by the available number of fast charging plugs and for even a low $EV_p=0.04\%$ it could reach 1 hour easily. 
The results highlighted in \cref{O1_O2_O3_O5_2016} indicate that current fast charging infrastructure deployed in our case study area is likely insufficient for even low levels of EV uptake and utilisation.  Even assuming that only $0.15\%$ of passing vehicles choose to recharge at a station, average waiting times balloon to unrealistic and unsustainable levels (exceeding 2 hours) in all but ten-plug stations.
Separately, note that total electricity consumption is highly sensitive to the placement of charging infrastructure. The average total energy consumption of 7 plug stations, which are located in high-volume traffic locations, far outstrip the consumption of the ten plug stations located in lower volume areas (if $0.15\%$ of passing vehicles charge, the 7 plug stations more than triple the energy consumption levels of the 10 plug stations).
 
The average service time (O3) is under 1 hour for the majority of stations regardless of the number of EVs waiting in the queue (see trend of \cref{O1_O2_O3_O5_2016}c)) while the energy consumption is highly dependent on the traffic flow recorded at specific stations and the utilisation of that station - for example in \cref{O1_O2_O3_O5_2016}d) the 7-plug stations are consuming almost 5 times more than even stations with ten plugs (at $EV_p=0.3\%$). The findings reveal not only the limitation of the current EV infrastructure in the area, but also allow to make predictions and planning decisions based on the popularity of specific areas in the network, their capacity, traffic demand, etc.

\subsection{EV impact on congestion}

\begin{figure}[ht]
	\centering
	\includegraphics[width=0.42\textwidth]{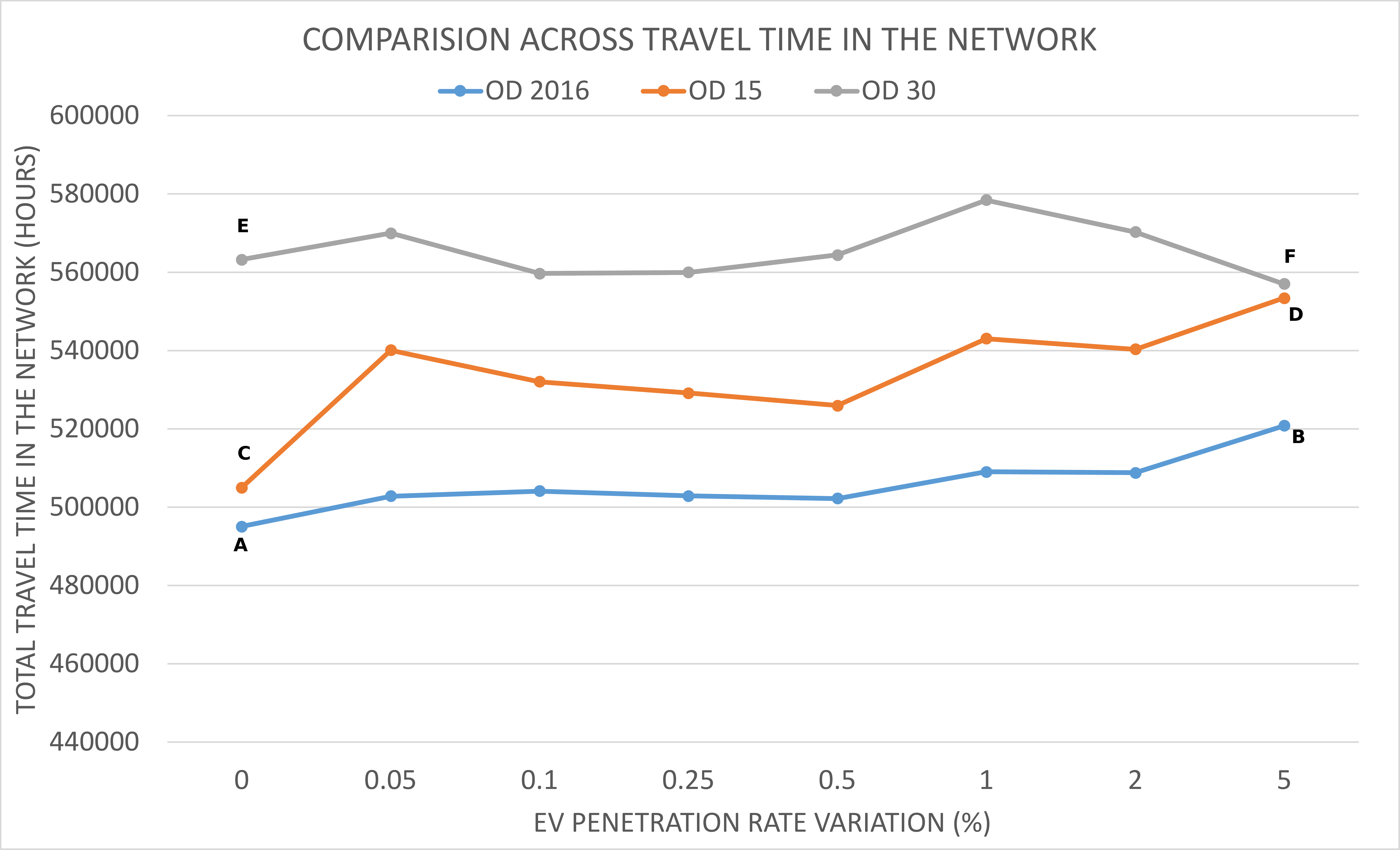}
	\caption{Travel time evolution across several traffic demand and EV penetration rates for OD2016, OD15, OD30 scenarios.}
	\label{Fig2_EV_impact_on_TravelTime}
\end{figure}

The mean queue results $\overline{Q^{S_i}}(T_r) [veh]$ of the EV-Q modelling is further utilised in the NSTM model to evaluate the impact that queue spill-back outside of EV stations would have on the overall traffic congestion in the network. \cref{Fig2_EV_impact_on_TravelTime} presents the daily total travel time spent by all cars in the network, across all EV penetration rates (Ox axis), for OD2016 (blue), OD15 (orange) and OD30 (grey). Assuming the traffic demand remains constant in 2016, the total travel time would increase by almost $4.9\%$ if the EV penetration rate would reach $5\%$ (see increase from point A to B). More specifically, this represents an extra of almost  $1074$ days spent in traffic by all vehicles (equivalent to an extra 6.18 hours spent in traffic for each driver - a highly penalising scenario). OD15 seems to add an even higher burden: an $8.7\%$ increase from point C to D translates in $11.6$ extra hours of time spent in traffic per year. The OD30 demand scenario translates in a $12.1\%$ increase assuming $EV_p=0.01$ remains the same, an unrealistic assumption given the global EV uptake trajectories (point E versus A). 

%


\subsection{EV impact on energy consumption}
 
\begin{figure}[ht]
 \centering
	\includegraphics[width=0.42\textwidth]{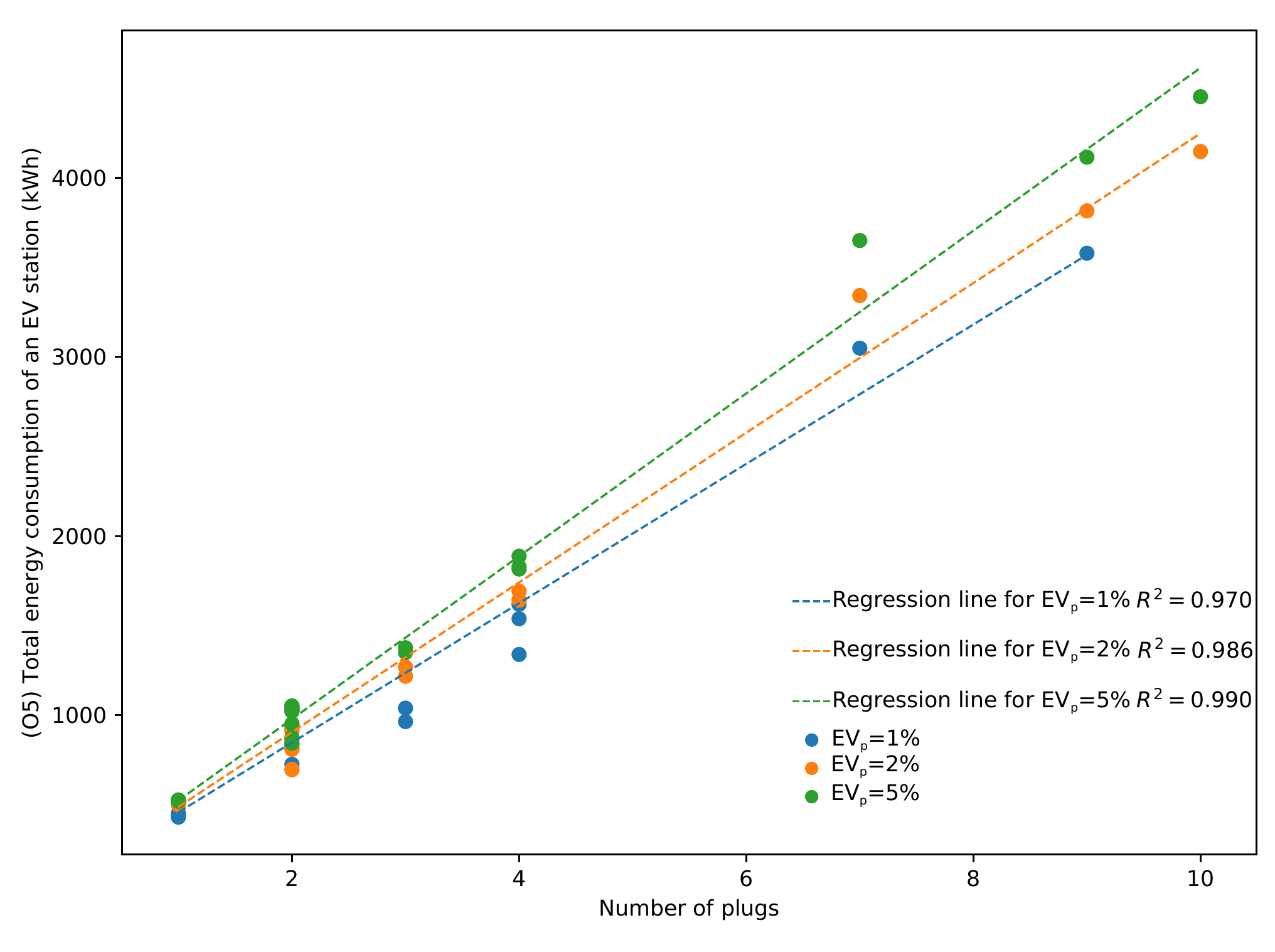}
	\caption{Energy consumption vs. the number of plugs for OD2016.} 
	\label{O5_vs_NOP2_small}
\end{figure}

While increasing the number of plugs can significantly reduce the waiting time for drivers, the energy impact will also increase considerably; \cref{O5_vs_NOP2_small} showcases the regression trends of daily total energy $E^{S_i}$(O5) versus $n_{ep}^{S_i}$; while stations with small $n_{ep}^{S_i}$ will not affect the energy consumption (for $n_{ep}^{S_i}<2$ and $EV_p=5\%$ the $E^{S_i}<1200kWh$), the stations with high number of plugs (7,9,10) will induce almost a 4-times higher energy consumption in the network ($E^{S_i}=4,500kWH$ for $EV_p=5\%$); this can have a  significant load on the electric network and the current EV station capacity.   

 \begin{figure}[h]
 \centering
	\includegraphics[width=0.4\textwidth]{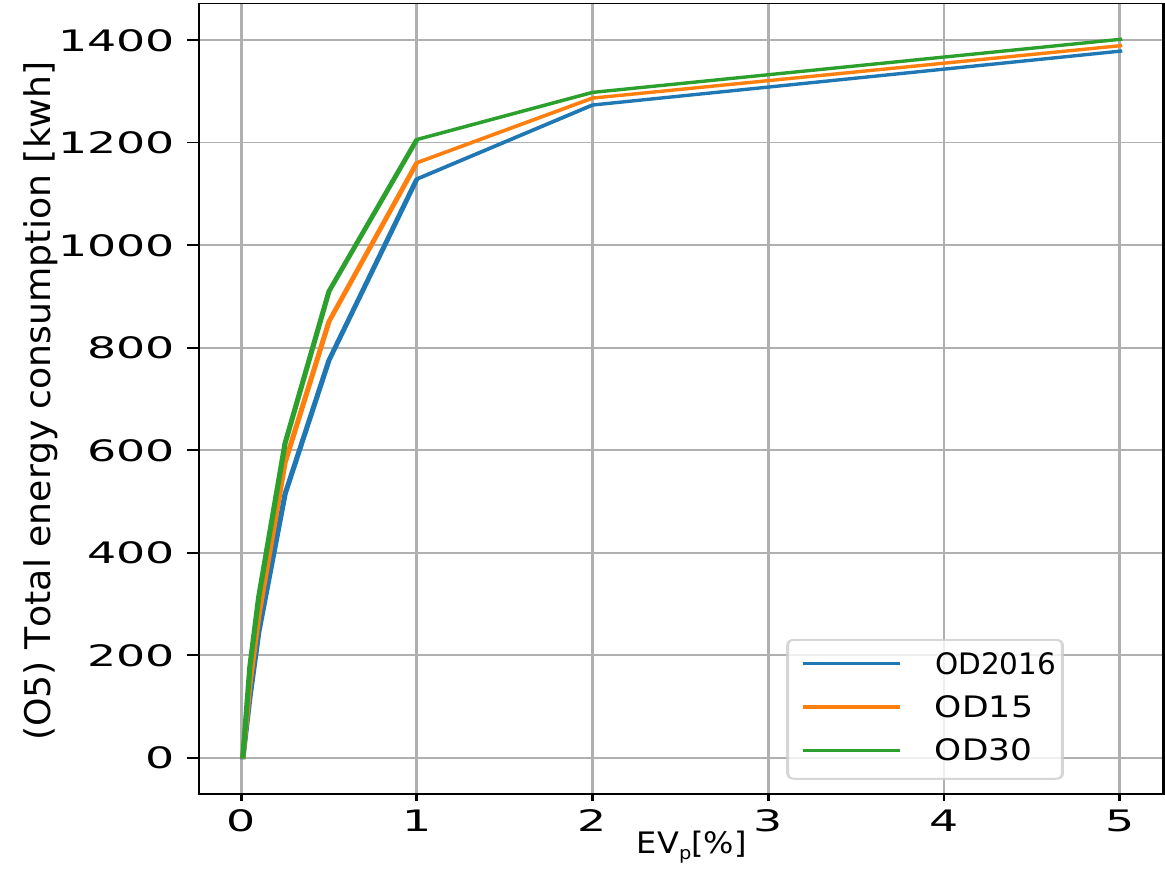}
	\caption{Energy consumption estimations for OD2016, OD15, OD30.} 
	\label{O5_5p}
\end{figure}
Furthermore, in \cref{O5_5p} we show that our demand increase assumptions for OD15 and OD30 translates in a saturation of the average daily energy consumption after an $EV_p=2\%$ (reaching 1400kWH) but follows similar trends across all traffic demands that we tested; this is mostly related to: a) the saturation of the traffic network when the demand increases considerably and vehicles would queue everywhere in the network and b) the EV infrastructure limitation which we assumed remains the same for our experiments despite the traffic demand would increase even by $30\%$ in the next years. There is however a significant impact on the energy consumption when switching from $EV_p=0.01\%$ to $EV_p=0.1\%$ which translates in an extra 1000kWh consumed daily on average across all stations. The impact of future EV increases and how this affects differently all EV stations, with an additional comparison is provided in~\cite{appendix}-\cref{EV_load} and \cref{TEC_Stations_2}. An important aspect is that EV stations with higher number of plugs (7,9,10) will be the ones that will increase almost three times more their energy consumption when compared to smaller stations (up to 4 plugs).

\subsection{EV impact on state grid}\label{Ev_grid_impact}
 \begin{figure}[h]
 \centering
	\includegraphics[width=0.49\textwidth]{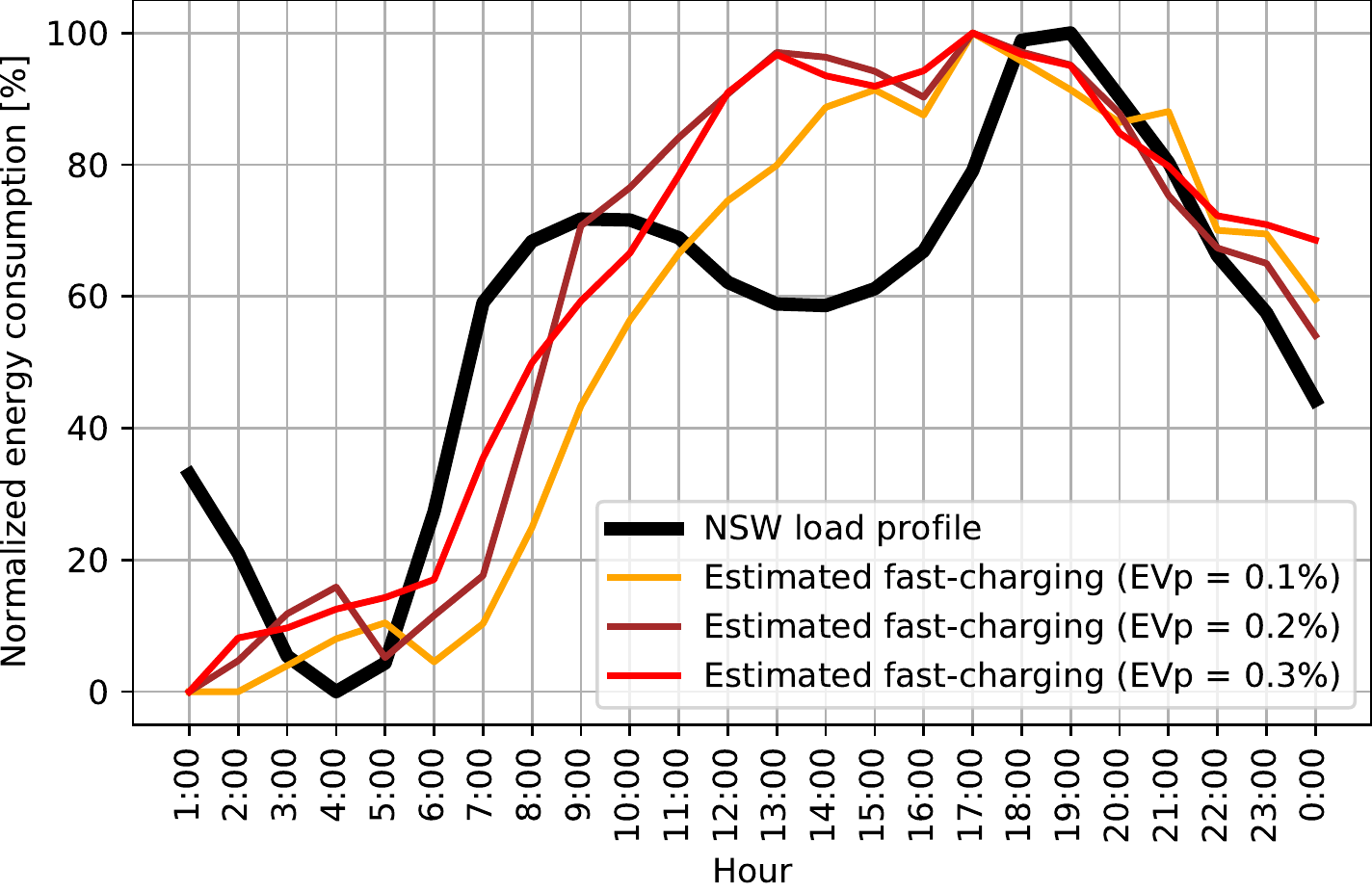}
	\caption{2016 state-level weekday energy consumption versus total estimated EV consumption across all stations in the EV-Q model.} 
	\label{EV_grid_impact_Adam}
\end{figure}
For analysing the EV impact on the electricity grid infrastructure, we compare the shape of aggregated EV charging demand with the shape of state-wide electricity demand in \cref{EV_grid_impact_Adam}. The results suggest that EV fast charging may exacerbate existing electricity peak demand issues.  Indeed, during the electricity peak demand period, modelled aggregate EV charging demand is between $91\%$ and $95\%$ of its daily peak. In many ways, the coincidence of electrical and charging demand is unsurprising: peak electrical demand in New South Wales is driven by residential energy use, particularly in the period following the end of the working day; traffic volume (and thus probable fast charger use) is similarly linked to working patterns and transit from work to home.

\section{Conclusion}\label{section_conclusion}
This paper presents an integrated bi-level framework of dynamic traffic modelling, data-driven queue and energy modelling with the purpose of evaluating the EV impact on both the traffic network and the energy consumption of the studies area. Results reveal existing limitations of the EV charging infrastructure which would not be able to handle even slight increases of EV rates, as well as significant travel time and waiting time degradation for larger EV penetration rates rates at each station level. The proposed bi-level framework was validated against real traffic and energy data sources and is currently being used for an extended analysis of this case study. 






%
%



{\footnotesize
\bibliographystyle{IEEEtran}
\bibliography{IEEE_ITSC_2021}}

\clearpage

  \appendix 

This document is accompanying the submission \textit{\titlename}.
The information in this document complements the submission, and it is presented here for completeness reasons. It is not required for understanding the main paper, nor for reproducing the results.
 
\subsection{NSTM model representation}

 \begin{figure}[h]
 \centering
	\includegraphics[width=0.49\textwidth]{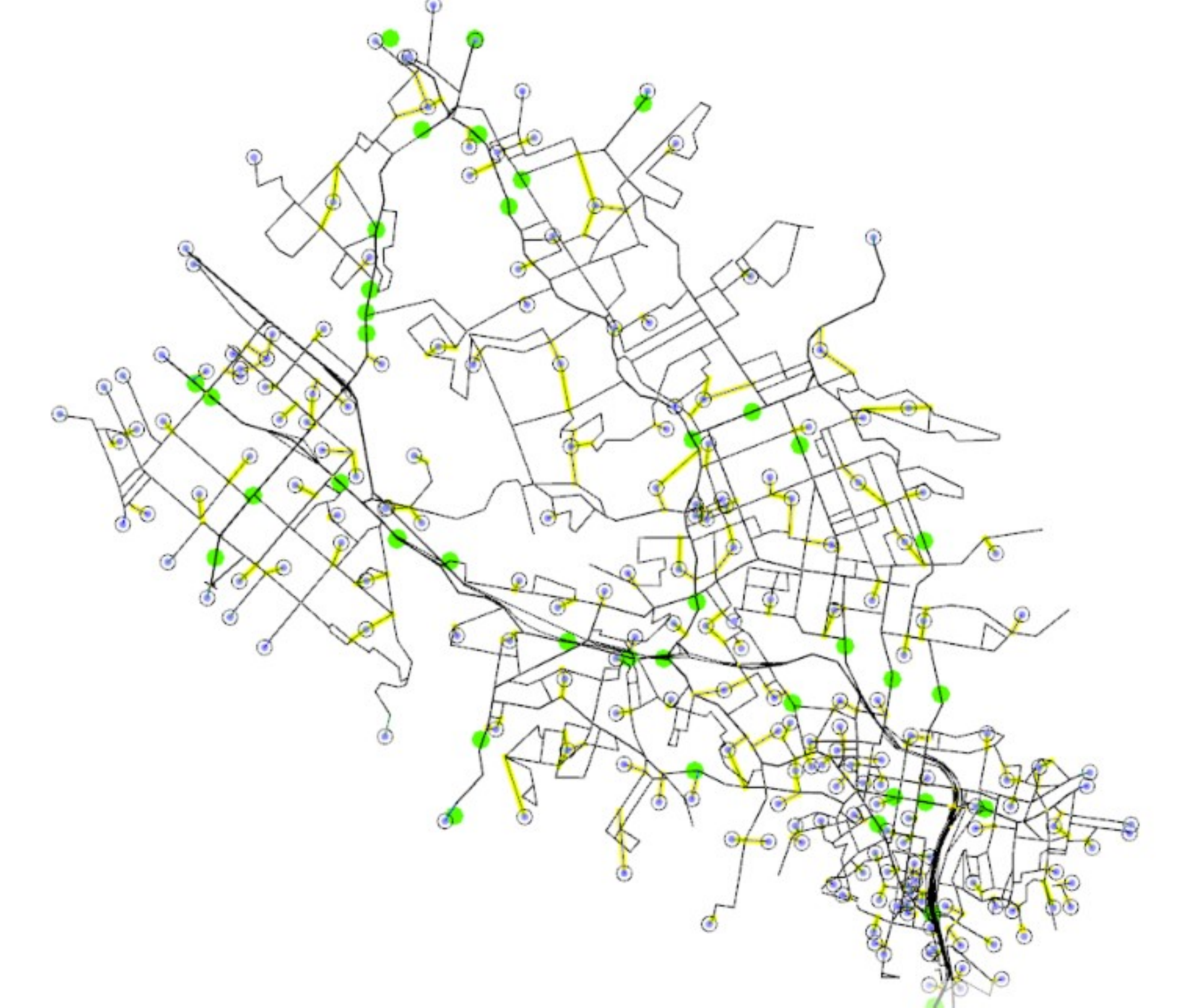}
	\caption{North Sydney Transport network layout including the connection of centroids (main entry points of vehicles in the network; marked as circles with blue interior) and permanent traffic count locations (marked in green dots).  }
	\label{Sydney_network_layout_EMME_Centroids}
\end{figure}

 \begin{figure}[h]
\centering
	\includegraphics[width=0.49\textwidth]{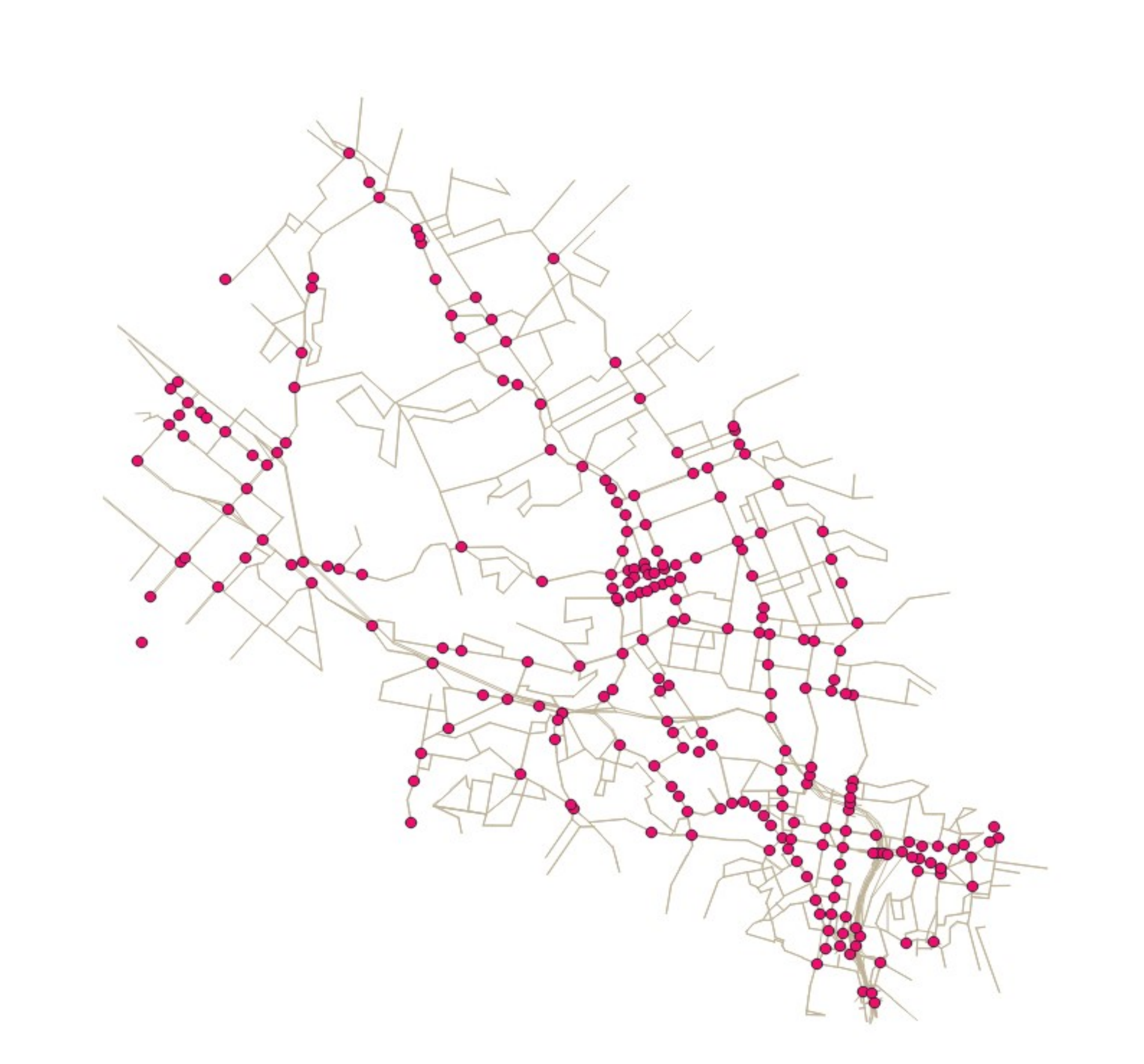}
	\caption[width=1\textwidth]{Layout of detector traffic flow.}
	\label{Sydney_network_layout_PTC_stations}
\end{figure}
 
 \begin{figure}[h]
\centering
	\includegraphics[width=0.4\textwidth]{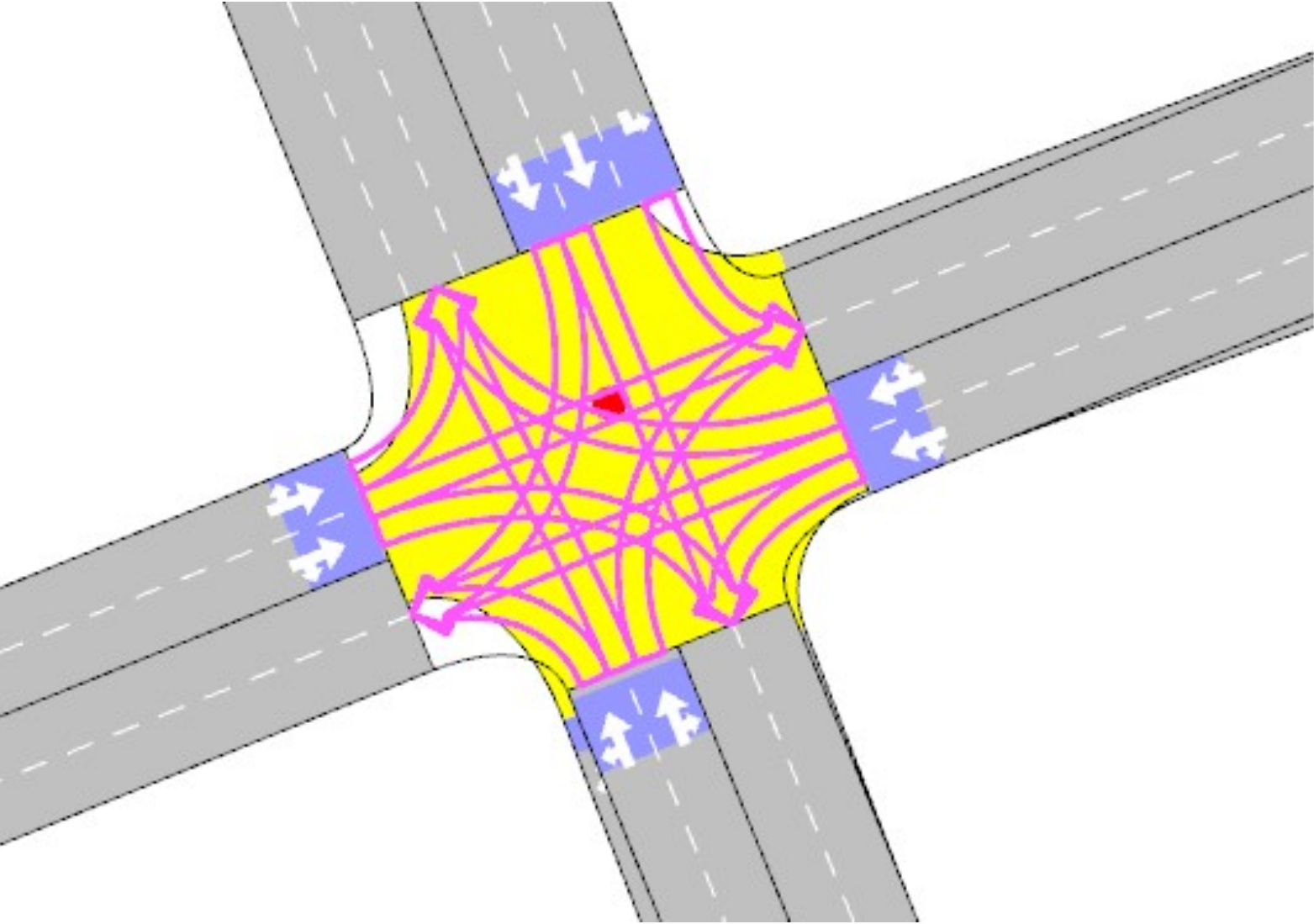}
	\caption[width=1\textwidth]{Sample representation of a signalised intersection including signal controls, phases, turnings and detectors for vehicle counting per each turning.}
	\label{SCATS_Control_intersection}
\end{figure}

\subsection{NSTM - PM peak validation}\label{PM_validation}

The model validation and calibration has been done as well for the PM peak hours as represented in figures below. 

\begin{figure}[h]
	\centering
		\includegraphics[width=0.4\textwidth]{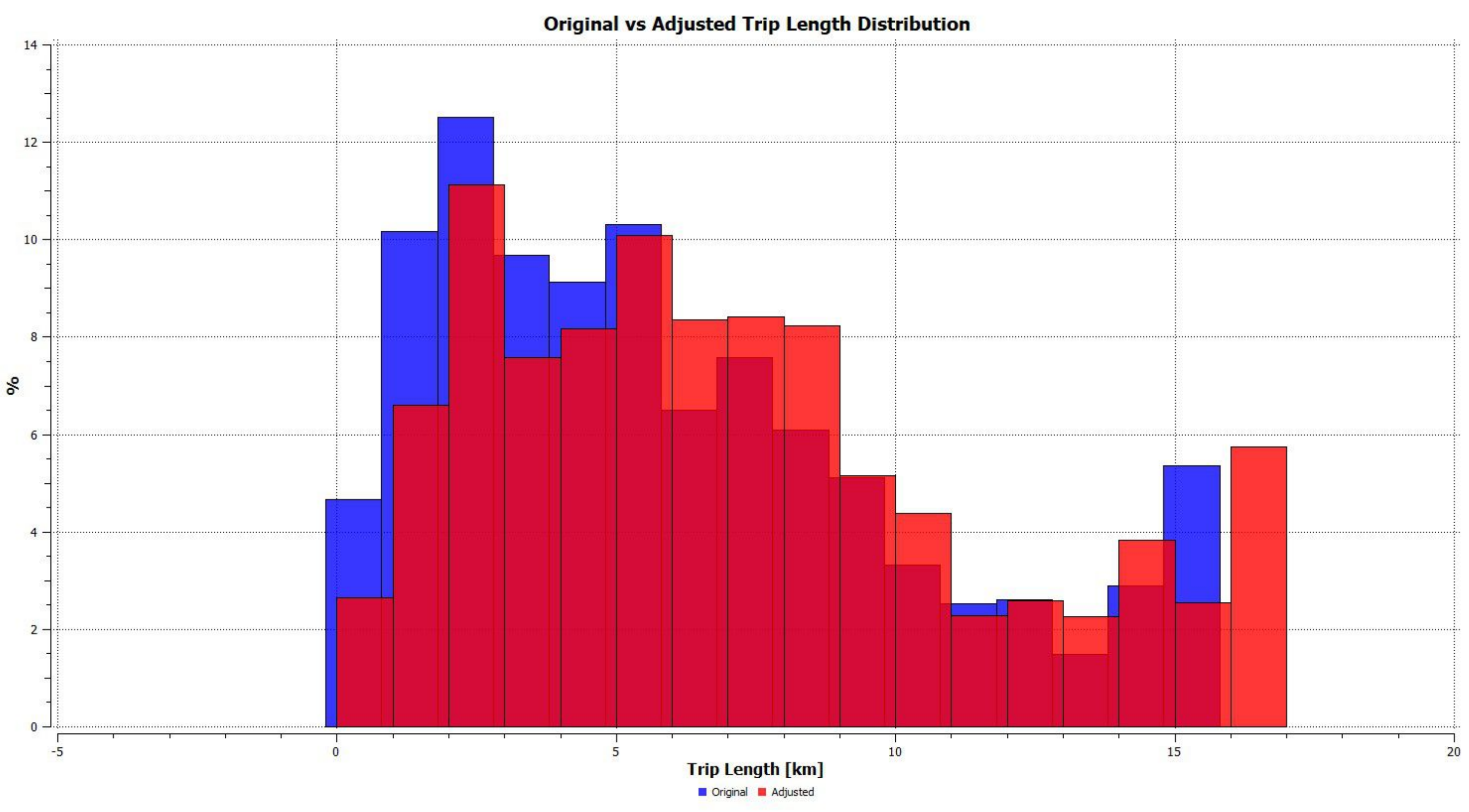}%
		\label{PM_OD_adjustment_fig}
\caption{Validation of PM-NSTM model via Trip Length Distribution }
\end{figure}
\begin{figure}[h]
		\includegraphics[width=0.4\textwidth]{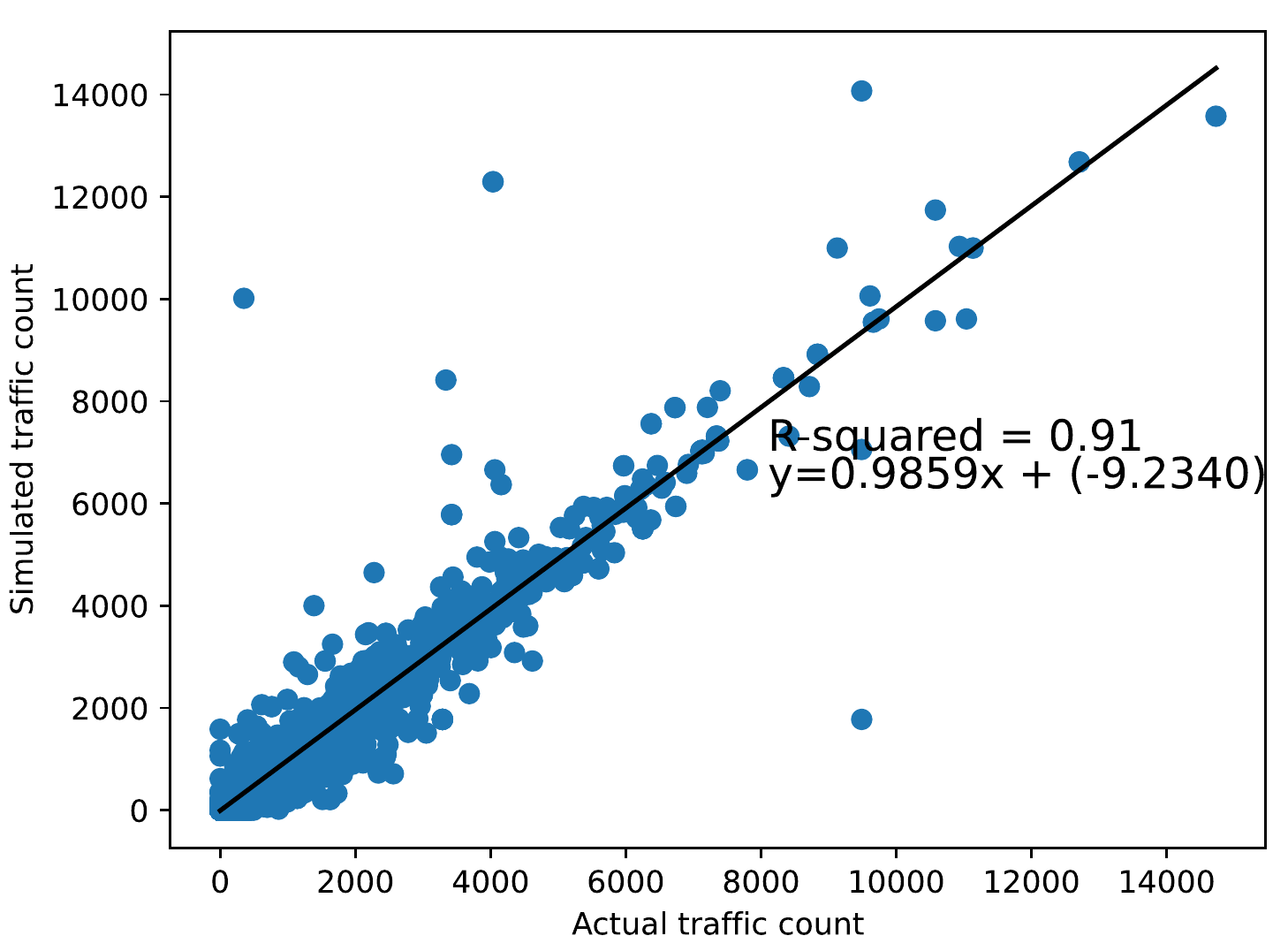}%
		\label{PM_R2_fig}
		\caption{Validation of the PM-NSTM model with $R^2=0.91$ metric.}
	\vspace{-.3cm}
\end{figure}
 
While the PM peak distribution after calibration indicates a lower number of adjusted vehicles travelling in the network, the trend profile still follows the original profiling of the data set.  In addition \cref{PM_R2_fig} indicates a high $R^2=0.91$ which is a proof of model well calibrated, despite few cases of over and under calibrated traffic counts.

\subsection{NSTM - car arrivals and mean distance travelled}\label{NSTM_distance_car_flow}
\begin{figure}[h]
 \centering
	\includegraphics[width=0.45\textwidth]{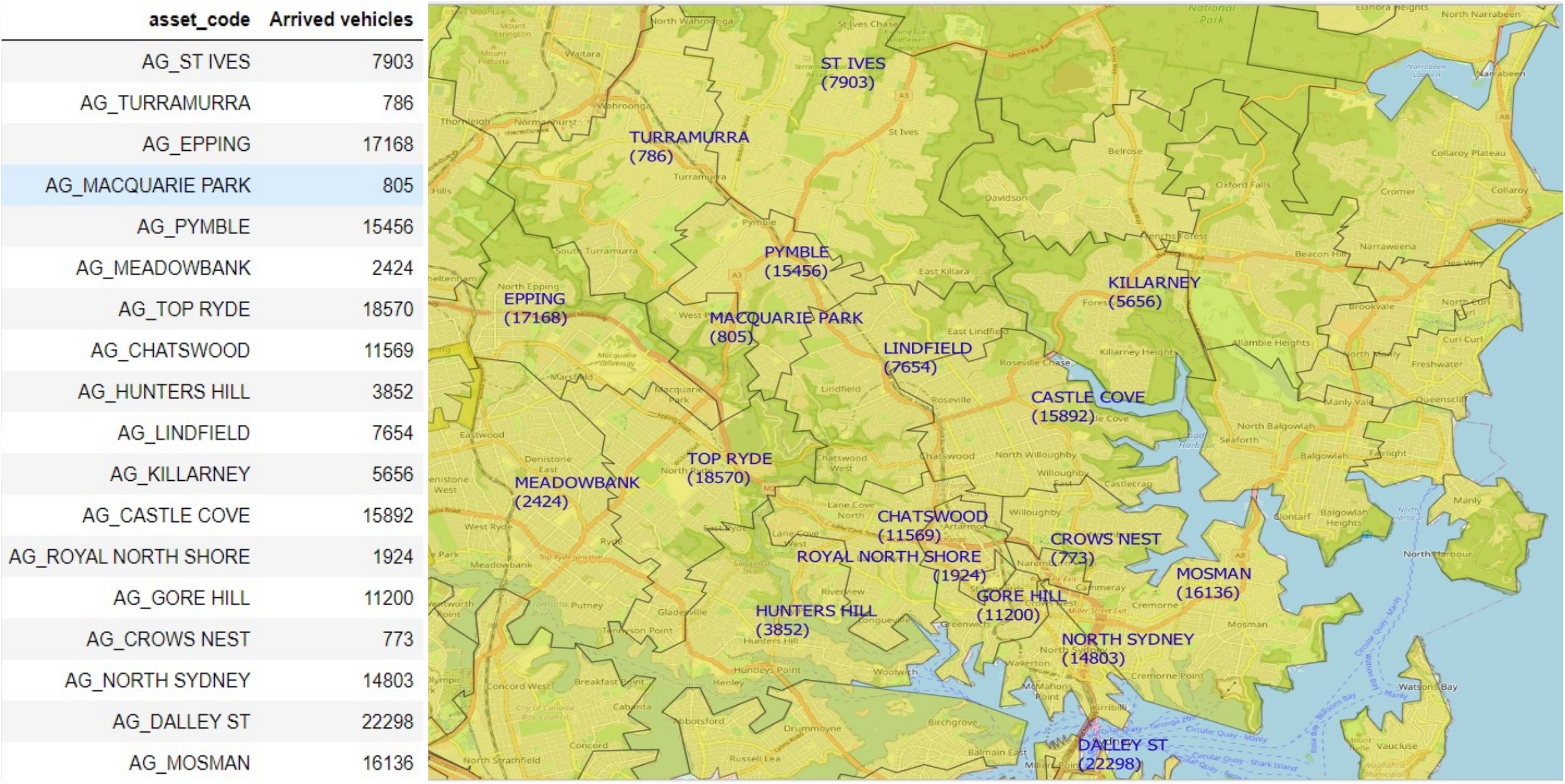}
	\caption{NTSM car arrival by area - AM peak.}
	\label{AM_car_arrivals_by_area}
\end{figure}

One of the outputs of the NSTM model is the total number of arrivals and departures from each area during AM and PM peaks. For modelling a home to destination charging, the number of arriving cars in each urban area is extremely important in order to be able to estimate the total number of EVs that will require charging at specific times. \cref{AM_car_arrivals_by_area} and \cref{PM_car_arrival_by_area} indicates the AM and PM car arrivals for each of the urban Local Governmental Area in our case study which allowed us to identify the top most attractive locations. For example, Dalley is a small catchment, but this area connects two parts of the city through Harbour Bridge. 
 \begin{figure}[h]
 \centering
	\includegraphics[width=0.4\textwidth]{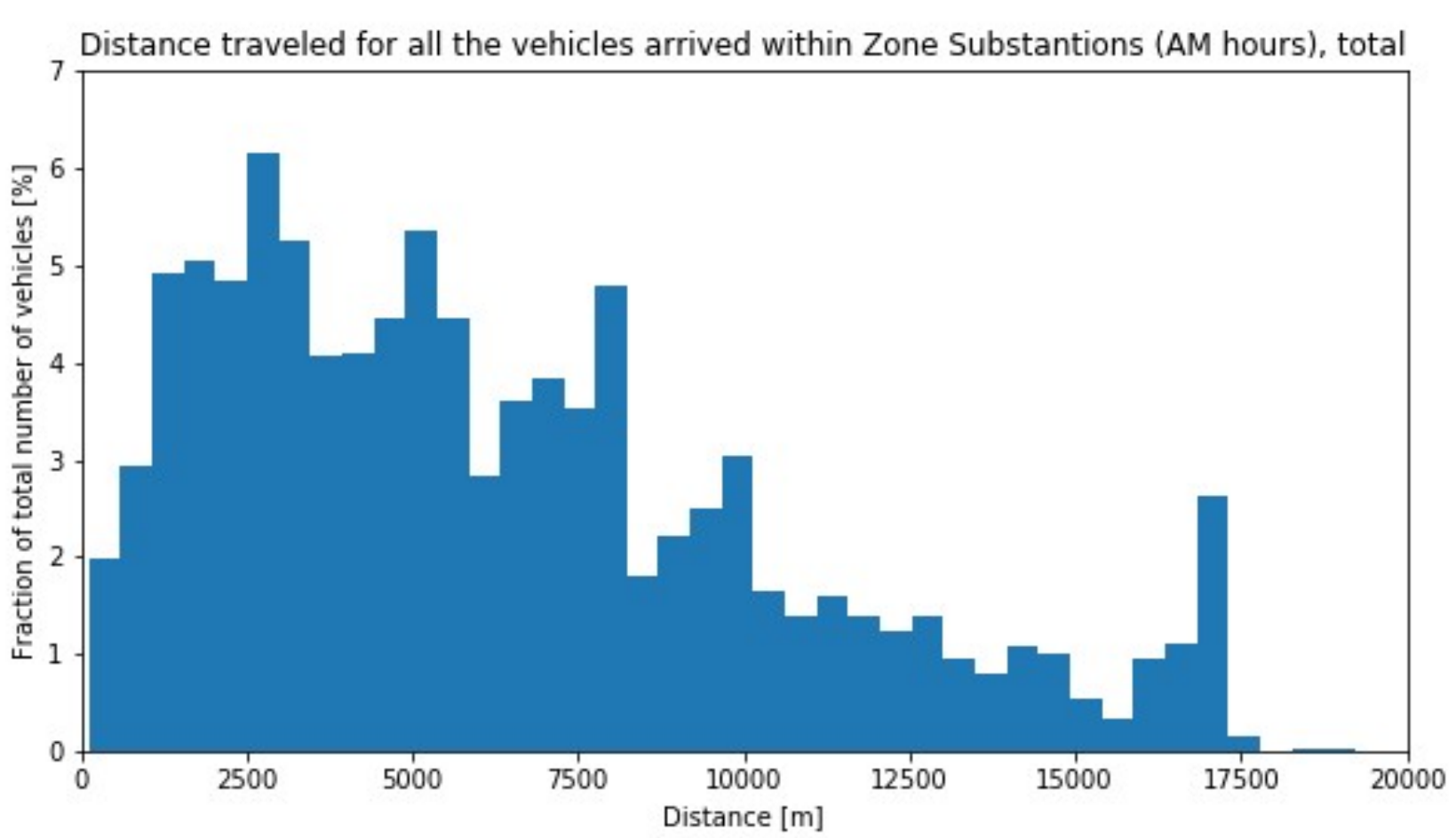}
	\caption{NTSM mean distance travelled - AM peak.}
	\label{AM_distance_travelled_by_area}
\end{figure} 
  \begin{figure}[h]
 \centering
	\includegraphics[width=0.45\textwidth]{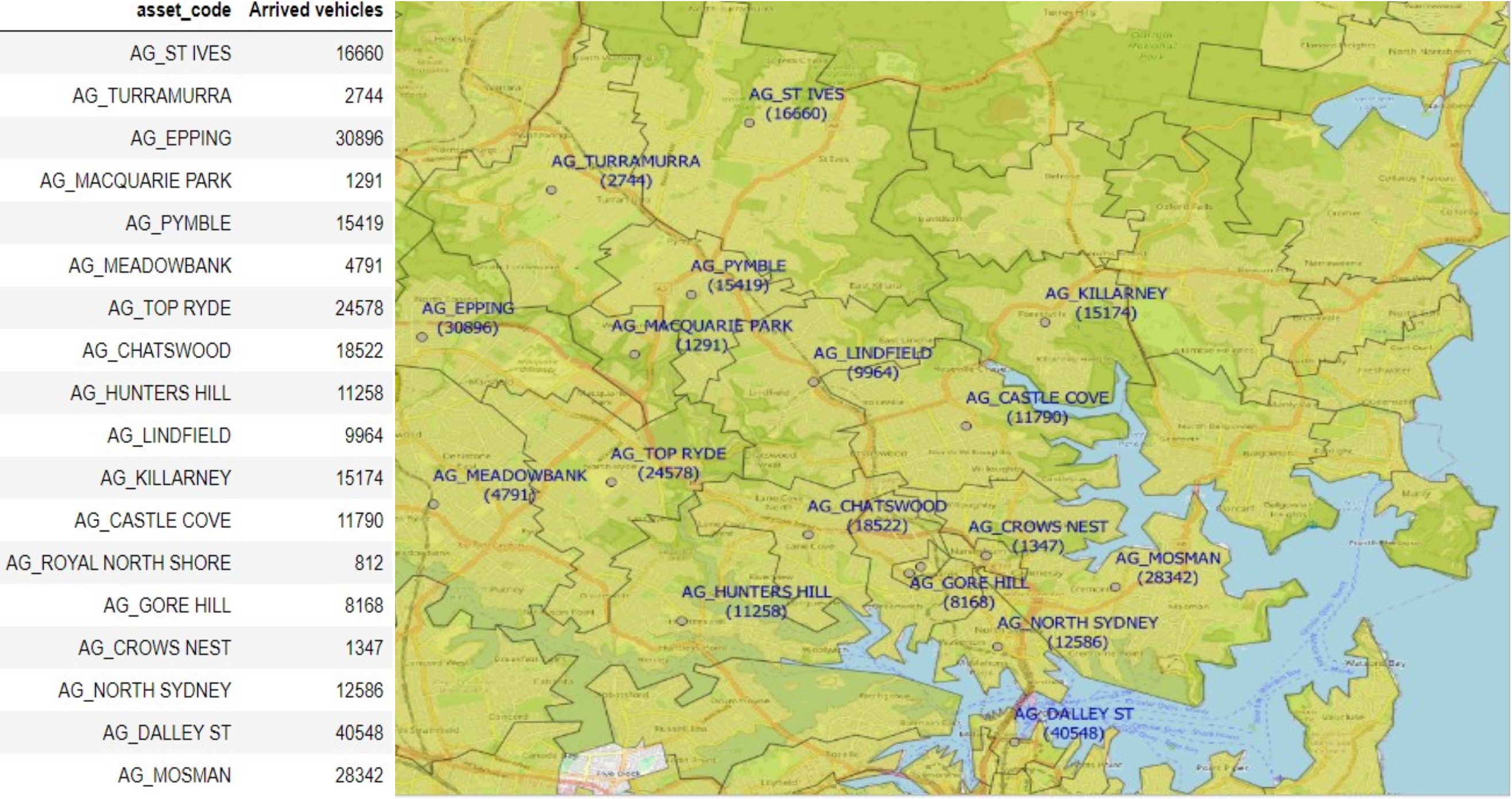}
	\caption{NTSM car arrival by area - PM peak.}
	\label{PM_car_arrival_by_area}
\end{figure}
 \begin{figure}[h]
 \centering
	\includegraphics[width=0.4\textwidth]{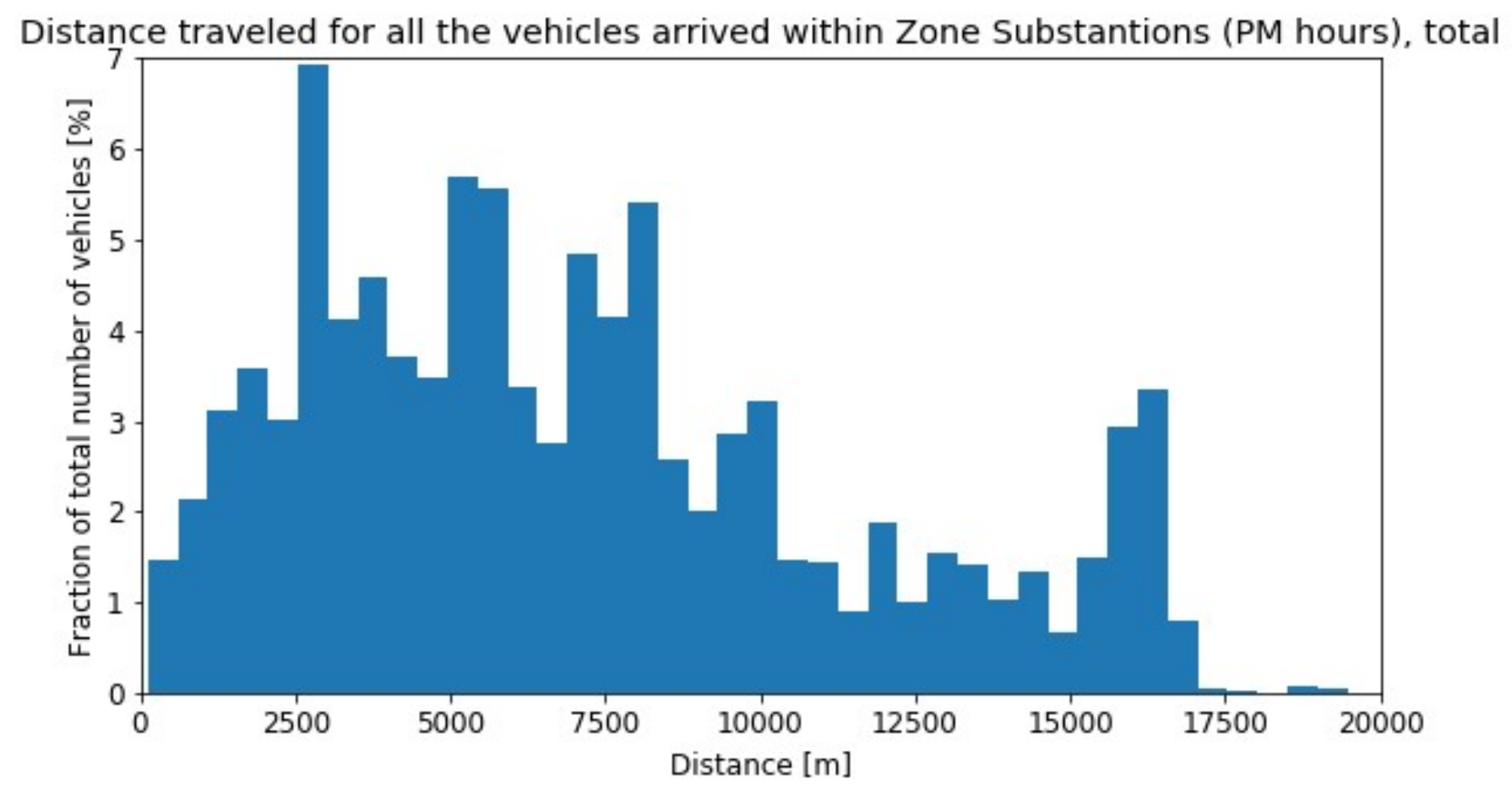}
	\caption{NTSM mean distance travelled - PM peak.}
	\label{PM_distance_travel_by_area}
\end{figure}
 \subsection{Capacity and loading of 10 most popular EV stations}\label{EV_load}
    \begin{figure}[h]
 \centering
	\includegraphics[width=0.4\textwidth]{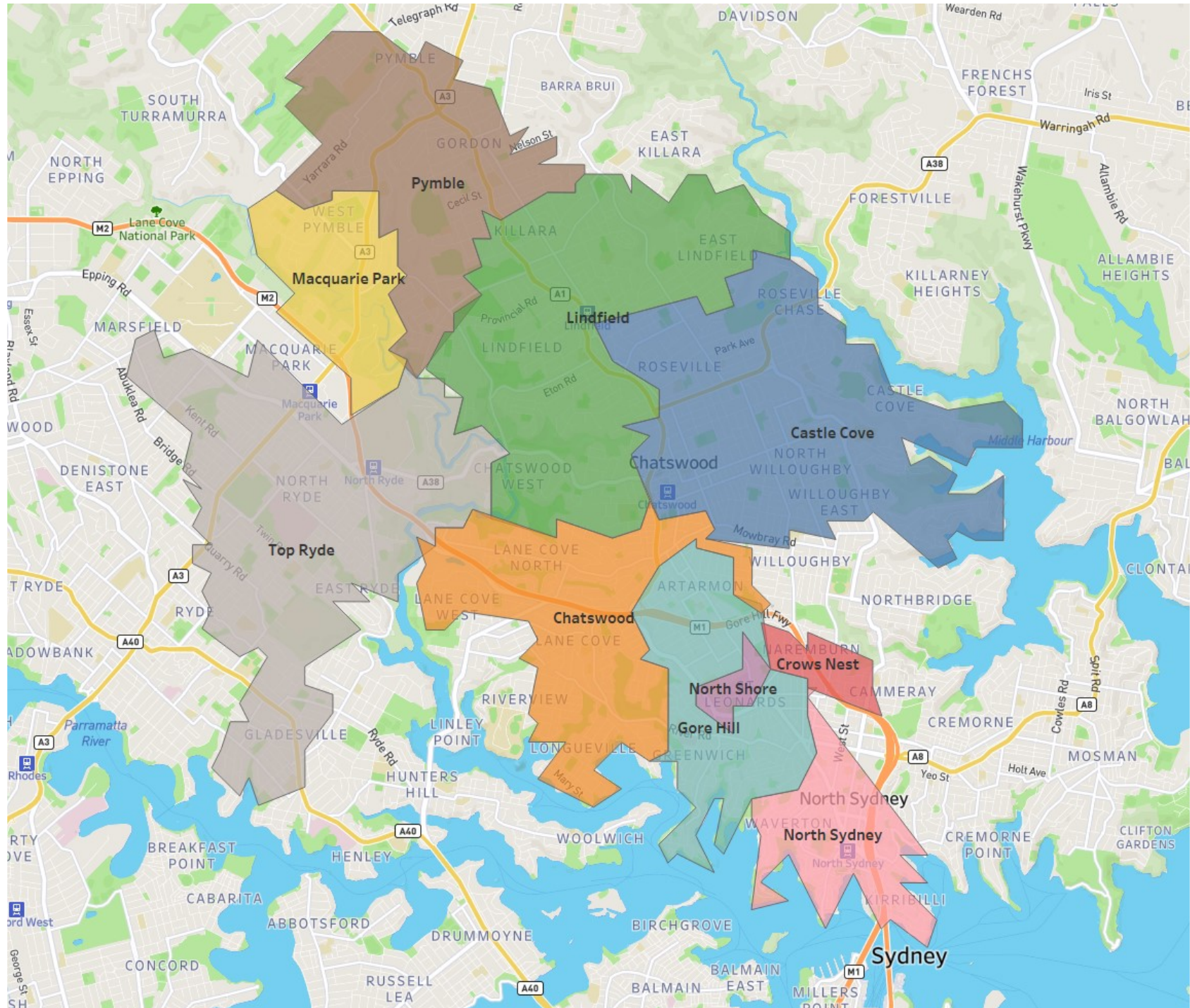}
	\caption{Zone substation area catchment.}
	\label{Zone_substation_area_selection}
\end{figure}
The Zone substation catchment area helped to identify the mapping of EV stations to the regional energy consumption and the analysis of the energy load for the top 10 most popular stations. Findings have revealed that 7 out of these 10 stations are operating at more than $75\%$ of their capacity, which already reveals current EV infrastructure limitation for further EV uptake. 
 \begin{figure}[h]
 \centering
	\includegraphics[width=0.4\textwidth]{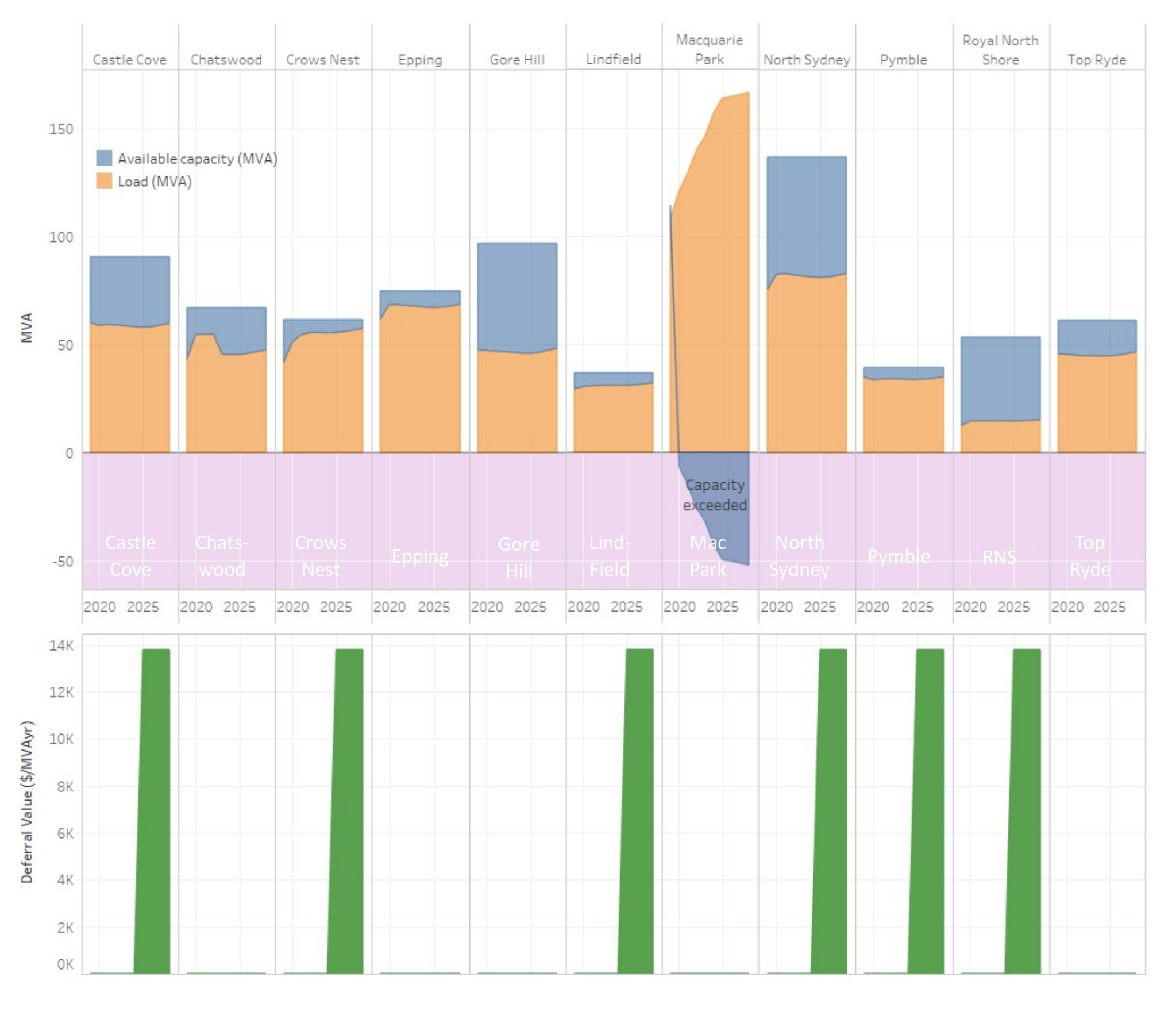}
	\caption{Load and capacity of 10 most popular EV stations.}
	\label{Traffic_flow_fluctuations_in_the_network}
\end{figure}

 \subsection{Energy impact of various scenarios}\label{EV_load}

 \begin{figure*}[ht]
 \centering
	\includegraphics[width=1\textwidth]{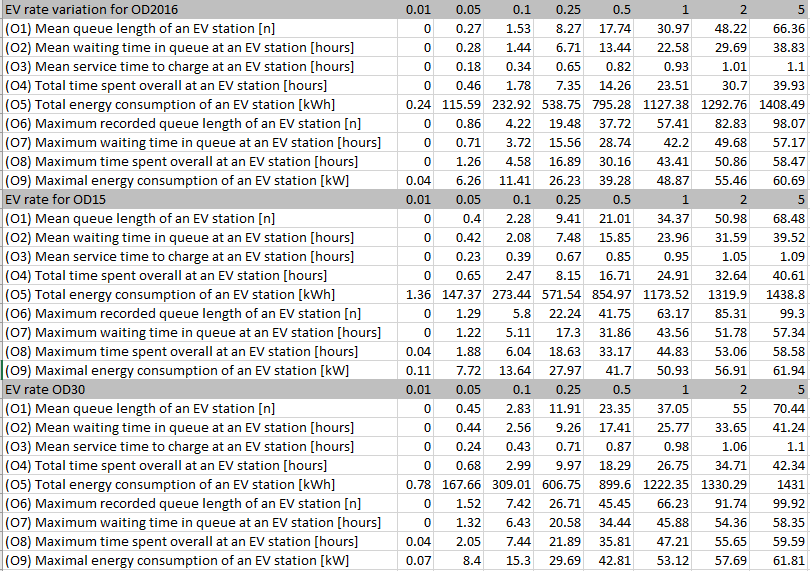}
	\caption{Raw results for EV-Q model outputs for OD2016, OD15, OD30 aggregated across all stations.}
	\label{2016_2026_2036_raw_results_table}
\end{figure*}

  \begin{figure}[h]
 \centering
	\includegraphics[width=0.45\textwidth]{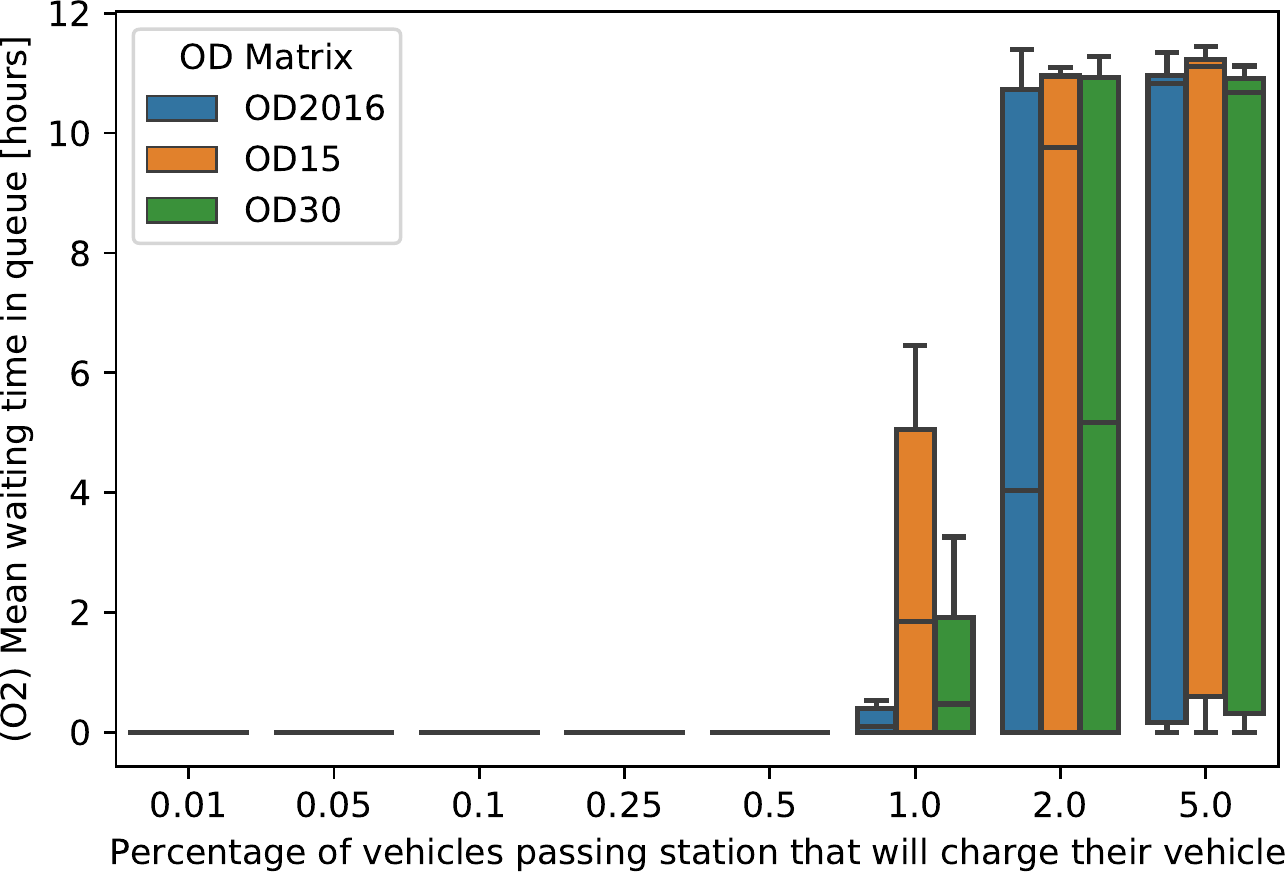}
	\caption{Mean queue variation for 10 plug EV stations.}
	\label{Queue_for_10_plug_stations}
\end{figure}

   \begin{figure}[h]
 \centering
	\includegraphics[width=0.45\textwidth]{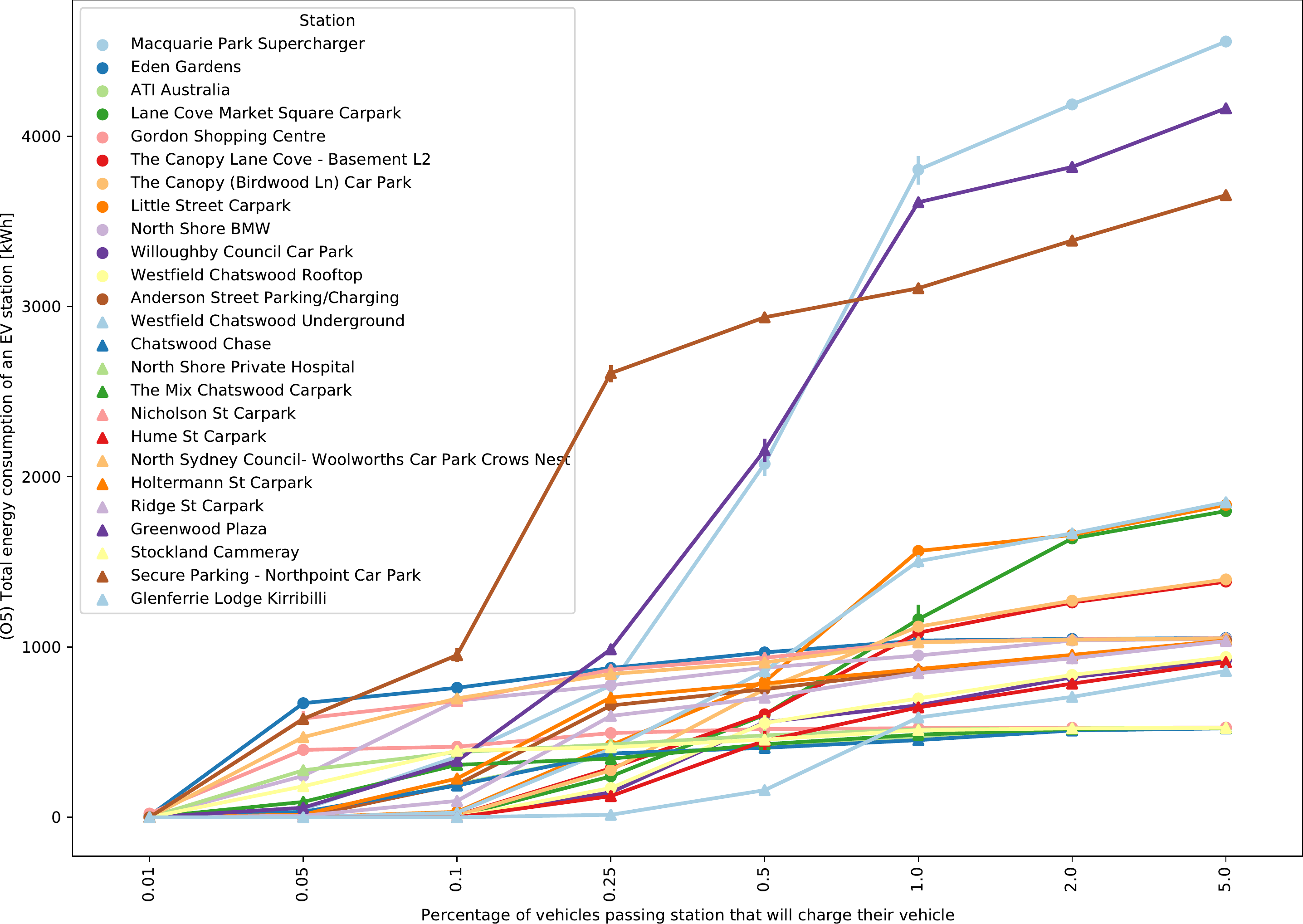}
	\caption{Energy consumption of all stations vs $EV_p$,}
	\label{TEC_Stations_2}
\end{figure}

\end{document}